\documentclass[11pt,titlepage]{article}
\usepackage{a4wide}
\usepackage{cell,citeparens,cite}
\usepackage[usenames,dvipsnames]{color}
\usepackage{amsmath,amssymb,subeqnarray,varioref}
\usepackage{amsmath}
\usepackage{authblk}
\usepackage{float,graphicx}
\usepackage{natbib}
\usepackage{caption}
\DeclareCaptionLabelFormat{adja-page}{\hrulefill\\#1 #2 \emph{(previous page)}}
\DeclareCaptionFormat{myformat}{\textbf{#1}#2#3\hrulefill}
\usepackage[font={sf,small}]{caption}
\restylefloat{figure}
\usepackage{gensymb}
\usepackage{parskip}
\usepackage{ulem}
\usepackage{tabulary}
\usepackage{pdfpages}
\usepackage{mhchem}
\usepackage{longtable}
\renewcommand\thefootnote{\fnsymbol{footnote}}
\newcommand*\samethanks[1][\value{footnote}]{\footnotemark[#1]}
\newcommand{\manuallabel}[2]{\def\@currentlabel{#2}\label{#1}}

\begin{document}


# LET-99-dependent spatial restriction of active force generators makes spindle's position robust.

H. Bouvrais,[1,2,*] L. Chesneau,[1,2] S. Pastezeur,[1,2] M. Delattre,[3] and J. Pécréaux[1,2,*]

**Running title:**
Dual control of pulling forces

1

**Abstract**

During the asymmetric division of the *Caenorhabditis elegans* nematode zygote, the polarity cues distribution and daughter cell fates depend on the correct positioning of the mitotic spindle, which results from both centering and cortical pulling forces. Revealed by anaphase spindle rocking, these pulling forces are regulated by the force generator dynamics, which are in turn consequent of mitotic progression. We found a novel, additional, regulation of these forces by the spindle position. It controls astral microtubule availability at the cortex, on which the *active* force generators can pull. Importantly, this positional control relies on the polarity dependent LET-99 cortical band, which restricts or concentrates generators to a posterior crescent. After delaying anaphase onset, we detected this positional pulling force regulation in *C. elegans* as a precocious spindle rocking with respect to anaphase onset. We ascribed this control to the microtubule dynamics at the cortex. Indeed, in mapping the cortical contacts, we found a correlation between the centrosome–cortex distance and the microtubule contact density. In turn, it modulates pulling force generator activity. We modelled this control, predicting and experimentally validating that the posterior crescent extent controlled where the anaphase oscillations started, in addition to mitotic progression. We found in particular that the oscillation onset position resists changes in cellular geometry and moderate variations of active force generator count. Finally, we propose that spatially restricting force generator to a posterior crescent sets the spindle's final position, reflecting polarity through the LET-99 dependent restriction of force generators to a posterior crescent. This regulation superimposes that of force generator processivity. This novel control confers a low dependence on microtubule and active force generator exact numbers or dynamics, provided that they exceed the threshold needed for posterior displacement. Interestingly, this robustness originates in cell mechanics rather than biochemical networks.

# INTRODUCTION

Asymmetric cell divisions, with their differing daughter cell sizes, contents, and fates, are essential to the development of multicellular organisms (1, 2). In the nematode *Caenorhabditis elegans*, as with many other species, the mitotic spindle is oriented along the polarity axis and displaced out of the cell centre (3, 4), and in turn helps position the cytokinesis cleavage furrow (5-8). Most asymmetric divisions include pulling forces from the cell cortex that are exerted on the astral microtubule plus ends, and these forces are key in positioning and orienting the spindle (3, 4, 9). In the one-cell nematode embryo, cortical forces are generated by a well-conserved trimeric complex which pulls on astral microtubules. The complex is made up of a dynein/dynactin complex, a LIN-5, homolog of NuMA, and the G-protein regulators GPR-1/2, which are mammalian LGN homologs (10). In an asymmetric division, GPR-1/2 proteins reflect polarity cues (11) through their asymmetric locations (12, 13), increasing the number of *active* force generators in the posterior side (11, 14, 15). By active force generator, we mean a complex either currently pulling from the cortex or ready to do so upon meeting a microtubule (10, 16, 17).

A still open question is the mechanism connecting the cortical polarity cues to the forces that set the spindle's final position, and how this mechanism ensures a robust positioning. The LET-99 DEP-domain containing protein is an important part of this mechanism (18) and acts downstream to PAR polarity proteins (19). Indeed, a band of LET-99, spanning from 45% to 70% of the embryo length, is devoid of force generation (20). To account for how LET-99 contributes to positioning the spindle, two possible mechanisms were proposed: First, LET-99 can relocalise all posterior GPR-1/2 to the posterior crescent and so increase its concentration, and in turn create an imbalance in the number of active force generators (15). Second and not exclusively, LET-99 can act by inhibiting the GPR-1/2 in the 45%-70% band and only constrain where in the cortex forces can be generated, a so-called spatial restriction mechanism. The higher concentration of GPR-1/2 in posterior embryo half (12, 13) would then be caused by another mechanism, still linked to polarity. It is however difficult to reconcile the first possibility, concentrating active force generators, with still observing a posterior displacement of the spindle upon depleting LET-99 (20). Conversely, while the second possibility is attractive, the spatial restriction mechanism is still unknown. One proposed possibility was the contribution of the orientation of the pulled astral microtubules (21), in which LET-99 could inhibit forces oriented backward with respect to spindle's displacement; it nevertheless does not account for LET-99 depletion resulting in a spindle's anterior shift (18-20). Overall, how the LET-99 cortical band contributes to the spindle's final positioning is still to be understood. By a modelling approach, including cell geometry and microtubule roles, we here offer a mechanism, which acts through microtubule dynamics.

In our initial "tug-of-war" physical model, we focused on the spindle oscillation and posterior displacement. We found that, along the anaphase course, an embryo-wide decrease in the force generator detachment rate from astral microtubule accounted for the oscillation build-up and die-down and for the increase in the pulling force imbalance causing spindle posterior displacement (16). The suggested temporal regulation was reinforced by the proposed link between spindle displacement timing and the cell cycle (22). In that initial model, we assumed

that astral microtubules were abundant at the cortex during anaphase, and that the only limiting factor was the binding/unbinding dynamics of the force generators. Since then, Kozlowski et al. proposed a model that accounted for spindle oscillation in which microtubules have a limited access to the cortex (23). Furthermore, some studies showed that changes of microtubule dynamics, mild enough to prevent catastrophic phenotypes, resulted in alteration of spindle's final position (24, 25). These results call for re-examining the role of astral microtubules in spindle positioning. Indeed, we recently noticed that the posterior centrosome position at which oscillation starts, hereafter called oscillation onset position, was not synchronized with mitotic progression in the cousin species *C. briggsae* (13), indicating that some other mechanism(s) contributed to regulating the cortical pulling forces.

Since microtubules would be at the core of this novel regulation of cortical pulling forces by spindle's position, after validating that this control also exists in *C. elegans*, we directly observed their availability at the cortex. Combining this result with the posterior restriction of active force generators by the LET-99 band, we expanded the modelling of spindle rocking and positioning consequently. We modelled and validated the dual control, *temporal* and *positional*, of pulling forces and took advantage of this so-called full-expanded model to dig into the robustness of the spindle's final position. Interestingly, time simulations based on this model and our experiments suggest that spatial restriction of active force generators by LET-99 combined to limited microtubule contacts to the cortex, not only account for spindle's final positioning by LET-99 but also suggest that details of microtubule number and motor number/dynamics contributed more modestly, making the spindle's final position robust.

## MATERIALS AND METHODS

### Culturing *C. elegans*

*C. elegans* nematodes were cultured as described in (26), and dissected to obtain embryos. The strains were maintained at 25°C and imaged at 23°C, with the exception of *gpr-2* mutant, *such-1* mutant, *let-99(RNAi)* and *spd-2(RNAi)* and their controls, which were maintained at 18ºC/20°C and imaged at 18°C or 23 ºC. The strains were handled on nematode medium plates and fed with OP50 bacteria.

### Strains

TH65 *C. elegans* (*Ce*) YFP::TBA-2 (α-tubulin) (27) and ANA020 *C. briggsae* (*Cb*) GFP::β-tubulin strains with a microtubule fluorescent labelling were used as the controls for the "landing" assay. TH27 *C. elegans* GFP::TBG-1 (γ-tubulin) (28) and *C. briggsae* ANA022 (TBG-1::GFP;GFP::HIS-11) strains (13) displaying a centrosomal fluorescent labelling were the standards for the "centrosome-tracking" assay. For event timing, the control was the *C. elegans* TH231 (SPD-2::GFP) strain with centrosome labelling crossed to OD56 (mCherry::HIS-58) histone labelling. It was crossed with the KR4012 *such-1(h1960)* mutant strain (29) to create JEP16. Centrosome tracking upon mutating *gpr-2* was performed on the JEP14 strain, which was obtained by crossing the 10x backcrossed strain TH291 *gpr-2(ok1179)* and TH27 *C. elegans* GFP::TBG-1 (γ-tubulin).

### Gene inactivation through mutants or protein depletion through RNAi by feeding

RNAi experiments were performed by ingestion of transformed HT115 bacteria. *let-99* gene was amplified from AF16 genomic ADN and cloned into the L4440 plasmid. *cid-1* and *c27d9.1* were ordered from Geneservice. To obtain stronger phenotypes, the feeding was performed at 20°C for 48h (except for *let-99* and *spd-2*, which were only done for 16-24h and 6h, respectively). The control embryos for the RNAi experiments were treated with bacteria carrying the empty plasmid L4440.

**Preparation of the embryos for imaging**

Embryos were dissected in M9 buffer and mounted on a pad (2% w/v agarose, 0.6% w/v NaCl, 4% w/v sucrose) between a slide and a coverslip. Depending on the assay, they were observed using different microscopic setups. To confirm the absence of phototoxicity and photodamage, we checked for normal rates of subsequent divisions (30, 31). Fluorescent lines were imaged at 23°C unless otherwise indicated.

**Centrosome imaging**

For the "centrosome-tracking" and the "event-timing" assays, embryos were observed at the midplane using a Zeiss Axio Imager upright microscope modified for long-term time-lapse. First, extra anti-heat and UV filters were added to the mercury lamp light path. Secondly, to decrease the bleaching and obtain optimal excitation, we used an enhanced transmission 12 nm band pass excitation filter centred on 485 nm (AHF analysentechnik). We used a 100x/1.45 NA Oil plan-Apo objective. Images were acquired with an Andor iXon3 EMCCD 512x512 camera at 33 frames per second and using their Solis software. Images were then stored using omero software (32). Except if histones were labelled (Table 1), the beginning of the spindle's abrupt elongation (Figure S4) was used as the marker for anaphase onset (33), and the centrosome tracks of individual embryos were aligned using this reference, for averaging purposes or for overlay on the "landing" assay.

**"Landing" assay**

To measure the spatial distribution of microtubule cortical contact densities, we viewed the cortex of one-cell embryos with entire microtubule labelling through tubulin fluorescent tagging (Figure 1D), using a spinning disk microscope (LEICA DMI6000 / Yokogawa CSU-X1 M1) equipped with a HCX Plan Apo 100x/1.4 NA oil objective. Illumination was performed by a white-light Fianium laser filtered around 514 nm using an homemade setup (34), sold by Leukos (Limoges, France). We then denoised the images by Kalman filtering (35), and detected microtubule contacts by applying u-track (36). A detailed version of this analysis pipeline, together with the parameters used, could be found in the Supplementary Methods. The recovered lifetimes of the microtubule contacts followed an exponential distribution as expected from a first order stochastic process (Figure 1E) (23). To spatially map the microtubule contacts with a reasonable accuracy, for each embryo, we divided the cortex into ten regions of equal width along the anteroposterior (AP) axis and counted the number of contacts in each region along mitosis. The densities were computed by dividing these counts with the respective cortical region areas, which were precisely measured using the active contour algorithm (37). We then performed an averaging over 10 s blocks. The final density map was obtained by averaging over embryos after the density maps were aligned temporally according to the onset of cytokinesis furrow ingression. This temporal cue was set when embryo

shape decreases its convexity as reported by the ratio of convex area to active contour area (Figure S4B).

**Robustness plot (Figure 4)**

To assess the robustness of the position and timing of the posterior centrosomal oscillation onset against embryo length variations, we calculated dimensionless quantities. For the timing, we used the reference duration $T$ equals to the delay between two mitotic events independent of cell mechanics for control embryos. We chose the Nuclear Envelope BreakDown (NEBD) and the anaphase onset. For each experiment, we computed $t_0$, the shift of the oscillation onset time with respect to the anaphase onset. The normalized shift $\overline{\delta t_0}$ was thus obtained by subtracting the corresponding mean value for control $t_0^{ctrl}$ from the current value $t_0$, and by dividing the result by the reference duration: $\overline{\delta t_0} = (t_0 - t_0^{ctrl})/T$. We repeated this calculation for the positional quantities using the control's mean embryo length $L$ as a reference for normalization. For each experiment, we computed the shift of the position of oscillation onset, $p_0$, with respect to the corresponding control mean position $p_o^{ctrl}$. Normalization yielded the normalized shift of the oscillation onset $\overline{\delta p_0} = (p_o - p_o^{ctrl})/L$. Independently of the quantity used for normalizing, we used a Student's *t*-test to see whether the linear fit slope was significantly different from 0. In doing so, we were able to determine whether e.g. embryo length had an impact on the position or timing being studied.

**Simulation of the posterior displacement using the full-expanded model**

We simulated the posterior displacement of the spindle using our full-expanded model with the TR-BDF2 algorithm (38). To ensure the proper force balance on the spindle poles, we also included the anterior centrosome in the full-expanded model and restricted the anterior active force generators' location between 0% and 40% of AP axis (15). We restricted the anterior centrosome to a fixed position for the sake of simplicity, making the spindle position dependent only on the one of its posterior pole. It enabled us to compute the pulling force contribution coming from the anterior spindle pole. On the anterior side, we used a two-fold lower force generator on-rate (14), which results in half the number of *active* force generators compared to posterior side (15). We assumed that force applied to posterior centrosome and originated in anterior side was halved after anaphase onset because sister-chromatids separated. We modelled the centring force with a spring (33) and used the processivity to control the progression of mitosis (16). Finally, since the model was linearized, it was limited to considering modest variations in parameters around their nominal values. This simulation is further detailed in supplementary model.

**Statistics**

Averaged values were compared using the two-tailed Student's *t*-test with the Welch Satterthwaite correction for unequal variance except where otherwise stated. For the sake of simplicity, we recorded confidence levels using diamond or stars (◊, p ≤ 0.05; *, $p$ ≤ 0.01; **,

$p ≤ 0.001$; ***, $p ≤ 0.0001$; ****, $p ≤ 0.00001$) and n.s. (non-significant, $p > 0.05$; sometimes omitted to save room). We abbreviated standard deviation by SD, standard error by s.e., and standard error of the mean by s.e.m.

**Data processing, modelling, and simulation**

All data analysis was developed using Matlab (The MathWorks). Modelling was performed using Wolfram Mathematica formal calculus software. Numerical simulations were performed using Matlab and Simulink (The MathWorks).

# RESULTS

**Astral microtubule contacts at the cortex depend upon centrosome position in *C. elegans*, enabling the spindle to sense its position.**

To establish the nematode *C. elegans* as a suitable model, we first aimed to confirm that a positional control, *id est* a mechanism relating the cortical pulling force regulation with the spindle position, exists in this organism. Indeed, we previously reported that the position of the spindle's posterior pole controlled the onset of spindle oscillation in *C. briggsae* (13). In contrast with that nematode species, in which anaphase onset preceded oscillation by 30 s, in *C. elegans*, these onsets were simultaneous. This might be coincidental and we tested whether a positional control exists in this latter by delaying anaphase. We used a *such-1$^{ANAPC5}$(h1960)* mutant of the anaphase-promoting complex/cyclosome (APC/C) (29), labelling centrosomes and chromosomes using SPD-2$^{CEP192}$::GFP;HIS-58$^{H2B}$::mcherry. We tracked the centrosomes (16, 33), and observed oscillations that started largely before anaphase in the mutant (Table 1) but occurred when the posterior centrosome was at 70.8% of embryo length, like in the control (70.7%). In contrast, for both strains, the die-down of oscillation occurred about two minutes after anaphase onset, regardless of when oscillations started up, thus leading to different durations of oscillation phase (Table 1). We conclude that a positional control of anaphase oscillation onset exists in *C. elegans* embryos, similarly to *C. briggsae*.

Previous studies have emphasized the key role of microtubules in the positioning of the microtubule-organizing centre (MTOC). Indeed, they can "sense" cell geometry, for example to bring the MTOC to the cell centre (39, 40), or to orient the nucleus by exerting pulling forces that scale with microtubule length (41). To challenge our hypothesis that microtubule network provides a positional regulation of pulling forces, we decreased the microtubule nucleation rate through a mild *spd-2$^{CEP192}$(RNAi)* (27). By measuring the centrosome trajectories along cell division, we found that the posterior centrosome position at oscillation onset was significantly displaced posteriorly in SPD-2 depleted embryos, reaching 73.9% of embryo length compared to 70.2% in control (Table 1). We thus confirm that astral microtubules do play a role in regulating the centrosome's position at which oscillation starts.

| Characteristic measured (mean ± s.e.m.) | control (N = 17) | such-1(h1960) (N = 15) | control (N = 11) | spd-2 RNAi (N = 18) |
|---|---|---|---|---|
| Oscillation onset $T1$ (s) | -11.03 ± 5.50 | -39.83 ± 5.93 $p = 9 \times 10^{-4}$ | 9.36 ± 5.24 | 12.38 ± 5.37 $p = 0.66$ |
| Posterior centrosome (CS) reaching 70% of AP axis $T2$ (s) | -11.76 ± 5.68 | -85.50 ± 17.15 $p = 6 \times 10^{-4}$ | -0.54 ± 6.43 | -10.60 ± 5.19 $p = 0.074$ |
| Maximum oscillation amplitude (posterior) (%) | 20.78 ± 0.65 | 22.71 ± 0.61 $p = 0.035$ | 17.57 ± 2.22 | 17.76 ± 1.21 $p = 0.97$ |
| Nuclear envelope breakdown time (s) | -164.84 ± 3.48 | -291.04 ± 5.40 $p = 1 \times 10^{-17}$ | 229 ± 17 | 227 ± 10.4 $p = 0.91$ |
| Oscillation die down $T3$ (s) | 124.4 ± 7.0 | 117.5 ± 7.63 $p = 0.507$ | 170.6 ± 14.2 | 134 ± 5.3 $p = 0.12$ |
| Oscillation duration $T3 - T1$ (s) | 135.4 ± 8.9 | 157.3 ± 9.7 $p = 0.11$ | 166 ± 16 | 121 ± 7.6 $p = 0.15$ |
| Posterior centrosome position at oscillation onset (%) | 70.68 ± 0.89 | 70.76 ± 0.49 $p = 0.936$ | 70.25 ± 0.77 | 73.89 ± 0.77 $p = 0.001$ |
| Posterior centrosome position at oscillation die-down (%) | 79.51 ± 0.40 | 79.01 ± 0.69 $p = 0.515$ | 79.59 ± 0.35 | 80.41 ± 0.27 $p = 0.046$ |
| Embryo length (μm) | 52.60 ± 0.86 | 53.98 ± 0.80 $p = 0.237$ | 50.29 ± 0.95 | 50.37 ± 0.57 $p = 0.94$ |
| Embryo width (μm) | 35.21 ± 0.61 | 33.25 ± 0.53 $p = 0.019$ | 32.97 ± 0.54 | 31.97 ± 0.51 $p = 0.14$ |

**Table 1: Timing and position of metaphase and anaphase events in delayed-anaphase mutants and in reduced-microtubule-nucleation embryos.** Embryos with labelled centrosomes were viewed at 18°C and spindle poles tracked. We first compared labelled SPD-2[CEP192]::GFP;HIS-58[H2B]::mcherry embryos with (N = 17) and without (N = 15) non-null mutation such-1[ANAPC5](h1960). This gene codes for an APC/C (29). We also compared embryos, whose centrosomes were labelled using GFP::γ-tubulin, which were treated (N = 18) or not (N = 11) with spd-2(RNAi). Times were measured from the onset of anaphase, determined as the beginning of chromatid separation, when histone is labelled, or from the onset of spindle elongation (33). Peak-to-peak oscillation amplitude is shown as a percentage of embryo width. Positions along the AP axis are shown as a percentage of embryo length. Error bars indicate the standard error of the mean. $p$ values are reported for Student's $t$-test, Welch-Satterthwaite corrected.

Kozlowski and co-workers have reported that microtubules are not always abundant at any place at the cortex (23), we therefore hypothesized that the distribution of microtubule cortical contacts may depend on MTOC proximity, i.e. on centrosomes' position. We tested this by directly measuring microtubule distribution at the cortex ("landing" assay, Methods and Figure 1D). We used α-tubulin labelling to view both growing and shrinking states and did not observe any phenotype when comparing this strain with γ-tubulin::GFP centrosome-labelled strain (Figure S1AB, Methods). Because the microtubule dynamics are so fast, we viewed the cortical contacts at a speed of 10 frames per second. This labelling combined with the fast acquisition led to images with a low signal-to-noise ratio, which required a powerful denoising and adequate tracking (Methods, Figure 1A-C). We validated our image-processing analysis pipeline using simulated data (Figure S1C-E, Methods). We measured an histogram of microtubule contacting times at the cortex displaying an exponential distribution (Figure 1E), characteristic of a first-order process, with a lifetime consistent with previously published values (23). We then calculated the spatial distribution of the microtubule cortical contacts

along the AP axis. To reduce uncertainty, we block–averaged the distribution in ten regions of equal width along the AP axis (Figure 1D) and used a 10-second running average and averaged the result over the embryos. We observed spatial heterogeneity with two high-density ridgelines and an overall increase in contacts between metaphase and anaphase, the latter being consistent with the increasing nucleation rate previously described (27) (Figure 1F).

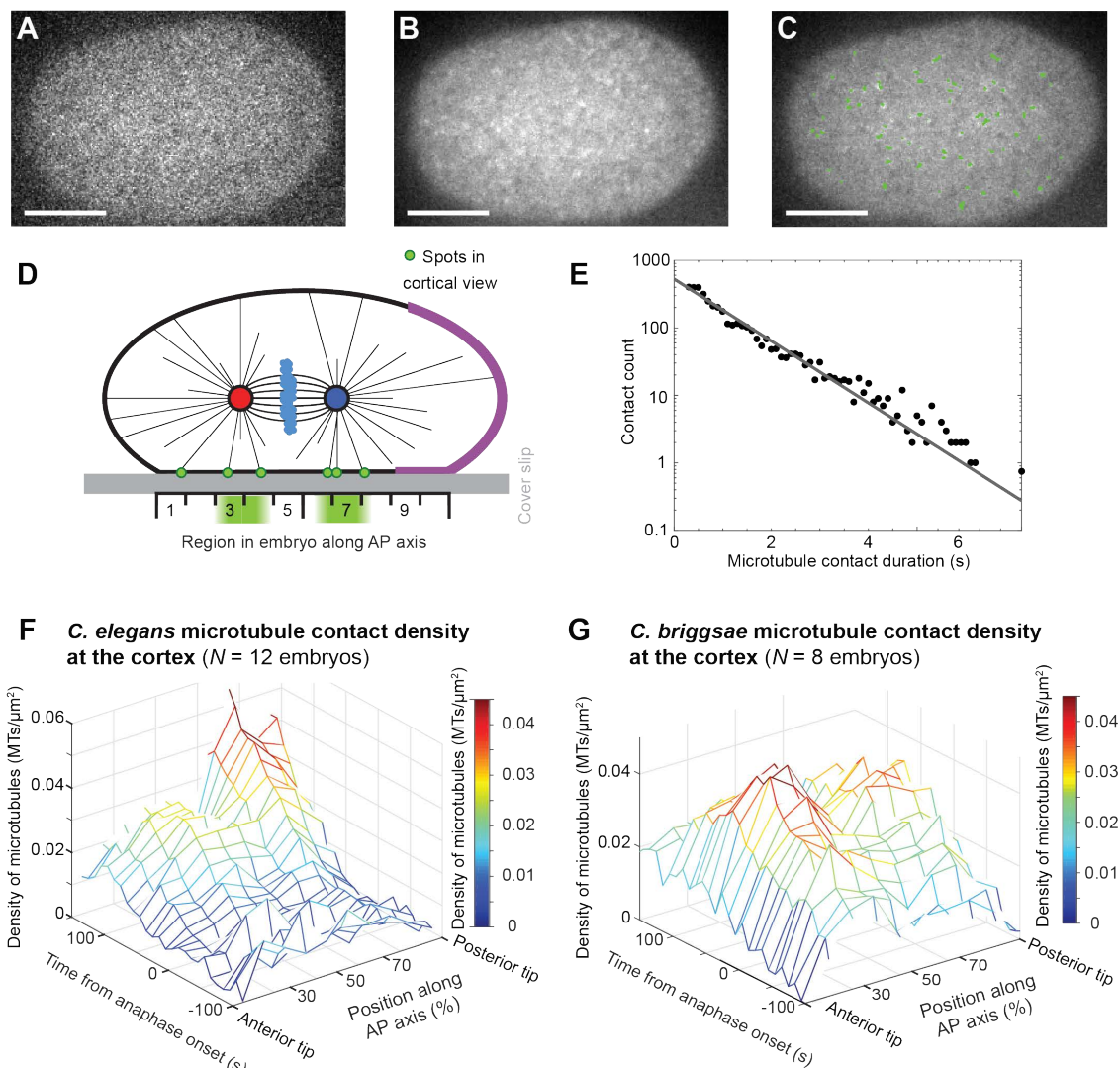

**Figure 1: Microtubule contact density at the cell cortex.**

(A-C) Exemplar spinning disk micrographs of *Caenorhabditis elegans* at cortical plane with YFP::α-tubulin labelling of the microtubules (MTs), viewed at 10 frames per second. Posterior tip of embryo is on the right side of the picture. The raw image (A) was denoised (B) using the Kalman filter and microtubule contacts tracked (C, green lines) using u-track algorithm with the parameters listed in Methods. Scale bars represent 10 μm. (D) Experimental setup for viewing microtubule contact density at the cell cortex. The scale represents the 10 regions along the anteroposterior (AP) axis used for analysis (Methods). Red and blue disks represent the anterior and posterior centrosomes, respectively, and the light blue clouds are the chromosomes. Microtubules emanating from the centrosomes are exemplified using thin black lines. The posterior crescent where the active force generators are located (so-called the active region) corresponds to the purple cortical region. (E) Semi-log plot of the histogram of the microtubule contact durations at the cortex during metaphase and anaphase for a single embryo (black dots), fitted with an exponential decay (grey line) corresponding to a characteristic time of 0.95 ±

0.03 s (*N* = 3832 microtubule contacts). (F-G) Microtubule contact densities at the cortex obtained by analysing spinning disk microscopy images of (F) *N=12 C. elegans* YFP::α-tubulin-labelled microtubule embryos and (G) *N* = 8 *Caenorhabditis briggsae* GFP::β-tubulin-labelled microtubule embryos. Microtubule contact densities measured at 23°C for each embryo were then averaged along the anteroposterior (AP) axis within 10 regions of equal width and over a 10 s running time window and finally the average over embryos was computed (Methods).

To test whether the ridgelines could correspond to the centrosomal positions, we used a wide-field microscope to view the spindle plane in the same strain and at the same temperature, and tracked the centrosomes. We then combined the results from both experiments and aligned them temporally with anaphase onset (Methods). We found that centrosome positioning coincides with the ridgelines (Figure 2A). Since we had initially observed the positional switch on cortical pulling forces in one-cell *C. briggsae* embryos (13), we thus wondered whether it could rely on a similar modulating of microtubule cortical contacts in space and time. We performed the same experiments in this species and obtained similar results (Figure 1G and 2D). We conclude that the distance of the centrosome to the cortex strongly controls the number of microtubules contacting the cortex in both species.

Only the force generators localized in the posterior crescent (so-called the active region) are able to exert a pulling force on an astral microtubule (20). We propose that the displacement of the area where microtubule contacts are concentrated toward the posterior along the course of mitosis could increase the number of microtubule contacts in the active region — a mean to regulate cortical forces. Could such a phenomenon create a positional switch? To address this question, we modelled the microtubule contacts at the cortex (Suppl. Model). Doing so, we included the dynamic instability of microtubules (42) and assumed that the force-dependence of the catastrophe rate was negligible (21). We also assumed that catastrophes happened only at the cortex (no free end catastrophe), and that microtubules fully depolymerized upon shrinking (negligible rescue rate) (23, 27, 43). We set a constant number of microtubule nucleation sites at the centrosomes, neglecting the modest increase in nucleation rate observed in anaphase compared to metaphase (27). Furthermore, these nucleation sites were never empty (21), and the microtubules emanated from there in an isotropic angular distribution (27, 44). We computed the number of microtubules that reached the cortex in the active region as a function of the position of the posterior centrosome (Figure 2B, black curve). This highlighted a limited number of microtubule cortical contacts when the centrosome position was close to cell centre with at first a slow increase in this number upon posterior displacement, followed by a steeper increase so that a high level was reached close to 70% of embryo length, consistent with the onset of oscillation observed at that position. To gain certainty, we sought to measure contacts in active region experimentally. We counted them in our "landing" assay and obtained a consistent measurement (Figure 2C). We propose that microtubule dynamics are at the core of the experimentally observed positional switch by regulating the number of microtubules available to force generators, making it dependent on centrosome position. Furthermore, the large number of microtubules that reach the active region during mid and late anaphase is consistent with the initial model's assumption that microtubules saturate a limited number of cortical force generators during this phase (16, 17). Overall, this paves the way towards

understanding the positional regulation of the cortical pulling forces involved in spindle rocking, posterior displacement and elongation.

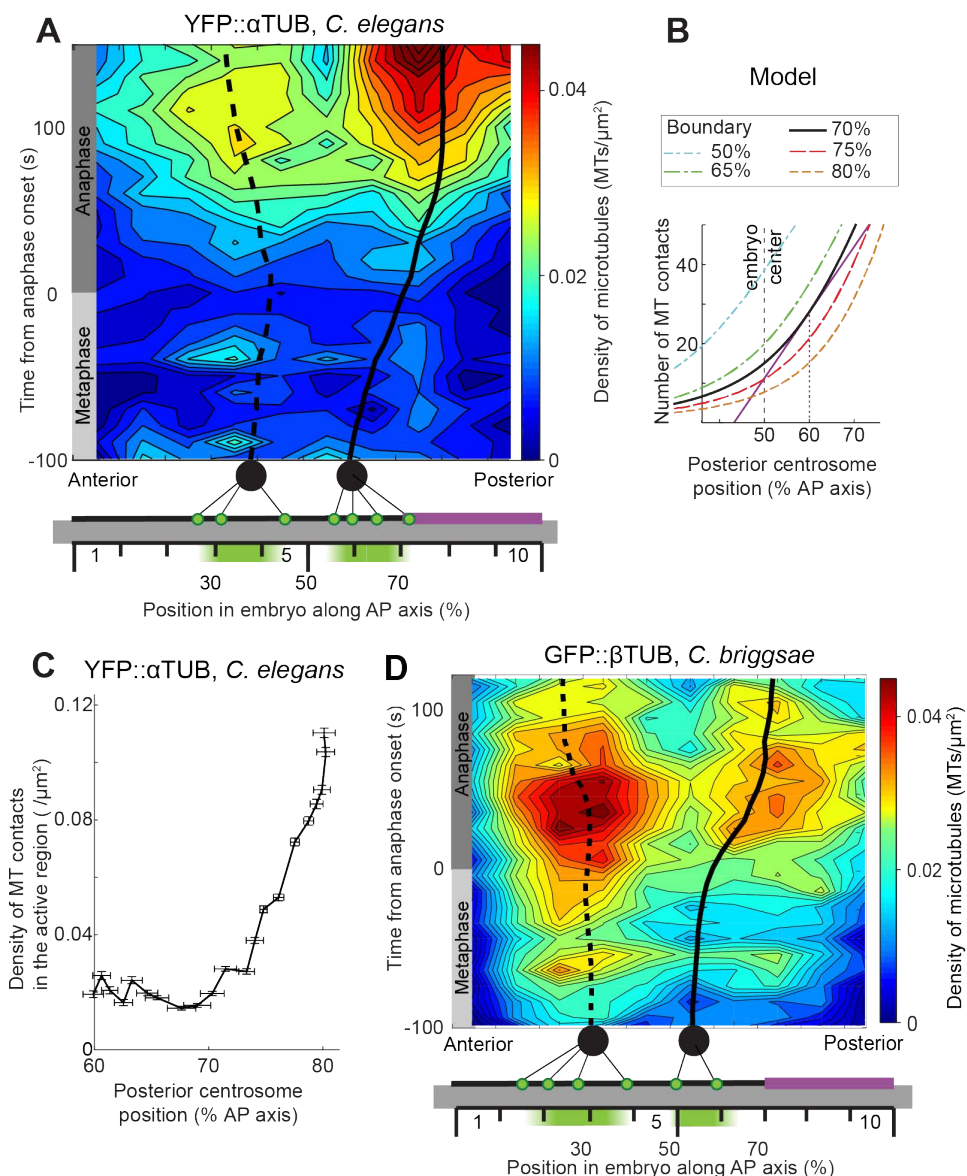

**Figure 2: Astral microtubules preferentially contact the cortex in the area closer to the centrosomes.**
(A) Microtubule (MT) contact density at the cell cortex in *Caenorhabditis elegans* with superposed centrosome trajectories. The densities, shown here as an interpolated heat map, were obtained by averaging the densities along the anteroposterior (AP) axis within 10 regions of equal width and over a 10 s running time window and finally taking the mean over embryos (Methods). Data are the same as in Figure 1F. At the bottom, schematic of the experimental setup where centrosomes are shown as black disks, with black thin lines depicting the astral microtubules. Their contacts at the cell cortex are shown as green dots. The active region, which is used to compute panel C, is depicted by the purple cortical horizontal line. The average trajectories of the centrosomes obtained by imaging the same strain in the spindle plane were superimposed ($N$ = 8 embryos). The dashed line represents the anterior centrosome trajectory, and the solid line indicates the posterior one. (B) Modelled number of microtubules contacting the cortex in the active region versus the posterior displacement of the centrosome along the AP axis, with the active region boundary expressed as a percentage of embryo

11

length. The thick black line corresponds to a boundary at 70%, mimicking the untreated embryo. In this case, the number of contacts started to increase steeply at a position above 60% (purple line). Blue and green curves model *let-99(RNAi)* experiments where the boundary was displaced anteriorly. Red and orange curves show the cases of posteriorly displaced boundaries. (C) Density of microtubule contacts *versus* the position of the posterior centrosome in a region ranging from 70% to 100% of AP axis, computed from the data shown in panel A. (D) Microtubule contact density at the cortex in *C. briggsae* with superposed centrosome trajectories represented similarly to panel A. Microtubule contact densities were measured at the cortex at 23°C (Methods). The data are the same as in Figure 1G. The centrosome trajectories were obtained by imaging the γ-tubulin::GFP;HIS-11::GFP-labelled centrosome and histone strain (*N* = 7 *C. briggsae* embryos) in the spindle plane and were then superimposed on the density map.

**Microtubule cortical contact modulation creates a positional control on the number of engaged force generators**

To quantitatively investigate how varying microtubule cortical contact density with spindle's position regulates cortical pulling forces, and to study later the impact on the spindle's final position, we further developed the physical modelling and extended our initial model (16). We propose that two modules combine to regulate the pulling forces (Figure 3A): a first one links the position of posterior centrosome to the number of microtubule cortical contacts in the active region which is in turn able to limit the number of engaged force generators, as modelled in the previous section; and a second module provides a temporal/mitotic progression control of the pulling forces through the dynamics of the force generators (16, 22, 45). Up to now, only the latter was accounted for in the initial model, and we envisioned that combining with the former would unravel novel robustness in the regulation of pulling forces and would enable to understand the mechanism linking cortical polarity and spindle's final position.

We created a first version of the two-modules model, termed "expanded model," where we still neglected the temporal evolution of force generator processivity (Figure 3A). We modelled the binding of microtubule and dynein as a first-order chemical reaction using the law of mass action assuming no cooperative binding between force generators (46) and estimated the association constant from the binding and unbinding rates used in the initial model (16). This enabled us to compute the number of engaged force generators versus the posterior centrosome's position (Figure 3B, black line). We found that when the centrosome was far from the posterior tip, we had a scarcity of astral microtubules contacting in the active region of the cortex. This limited the number of engaged force generators to below the previously described threshold for oscillating (16). Upon posterior displacement of the centrosome, past 60% of the AP axis, we observed a steep increase in engaged force generator count, similar to the one found in the number of microtubule cortical contacts (compare the black curves in Figures 2B and 3B). This was followed by a saturation starting from 70% of the AP axis. These two successive regime changes created a switch, which was consistent with our positional control hypothesis. In conclusion, our expanded model predicts that oscillation onset — a readout of pulling force increase — is regulated by the posterior displacement of the posterior centrosome.

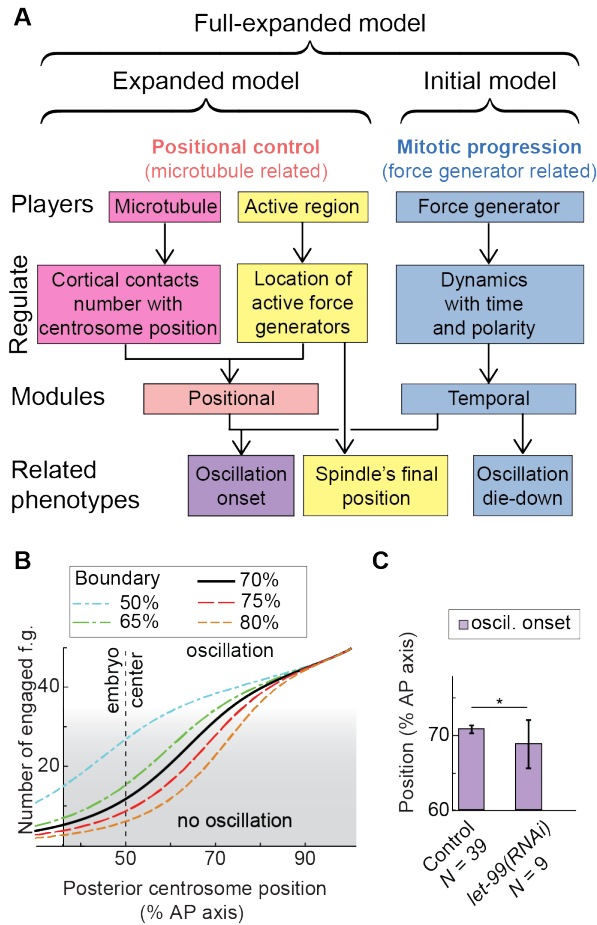

**Figure 3: Expanded model accounts for positional and temporal regulations of cortical force generators.**
(A) Schematics of the expanded model, highlighting the players (top row), the quantity regulated (second row), how they control forces (module, third row) and some related phenotypes (bottom row). Pink/yellow colours correspond to positional control, involving astral microtubule dynamics and the active region created by LET-99, and blue depicts time control, involving force generator dynamics. While both controls participate to oscillation onset (purple), spindle's final position mostly depends on active region extent (yellow), and oscillation die-down on time control (blue). (B) Modelled number of engaged force generators versus the posterior displacement of the centrosome along the anteroposterior (AP) axis, with the active region boundary expressed as a percentage of embryo length. The thick black line represents the case in which boundary is located at 70%, mimicking the untreated embryo. In this case, when the centrosome reached 60% of the AP axis, the number of engaged force generators increased steeply and saturated above 70%, causing a switch-like behaviour. Blue and green curves model *let-99(RNAi)* experiments where the boundary was displaced anteriorly. Red and orange curves show cases of posteriorly displaced boundaries. Grey shading indicates when the number of engaged force generators is too low to permit oscillation. The parameters used are listed in Table S5. (C) Position of the posterior centrosome at oscillation onset in *let-99(RNAi)* ($N$ = 9) and untreated ($N$ = 39) embryos at 23ºC, with centrosomes labelled by GFP::γ-tubulin. Error bars indicate SD, and star indicates significant differences (Methods).

We then challenged experimentally the expanded model in three ways related to each biological player (Figure 3A, top line) through measuring the timing of oscillation onset in respect to anaphase onset one, and the posterior centrosomal position when oscillation starts. Firstly, we

13

varied the extent of the active force generator region in the expanded model and could predict that the precise position at which oscillations started correlates with the boundary of the active region (Figure 3B). The extent of this region inversely reflects the LET-99 cortical band (20). We thus increased this extent towards the anterior side, still in the embryo's posterior half, by partially depleting LET-99 by *RNAi*. We observed that as compared to the control, the posterior centrosome's position at which oscillations began was significantly displaced towards the anterior (Figure 3C), in agreement with the expanded model's predictions (Figure 3B, blue and green curves). We conclude that the position of the active region boundary controls the position where oscillation onset occurs.

In a second challenging prediction, we focused on microtubules and predicted that a decrease in their total number emanating from a centrosome posteriorly displaced the position of oscillation onset (Figure S2A, blue and green curves). Experimentally, it corresponded to depleting the protein SPD-2 by *RNAi*, which decreased microtubule nucleation at the centrosomes (27), and resulted, as predicted, in a posteriorly shifted position of the posterior centrosome at oscillation onset (Table 1).

Thirdly, we varied the number of active force generators, which is predicted to have a reduced impact on the oscillation onset (Figure S2B, green curve), provided that it was above the needed threshold (16). We previously observed that a partial *gpr-1/2(RNAi)* did not significantly change the oscillation onset position (13). We set to reproduce this result in the present paper using a non-null *gpr-2(ok1179)* mutant at 18ºC. To estimate the remaining level of force generation, we observed posterior centrosome's oscillations and measured a reduction of their amplitude to $9.0 \pm 3.3\%$ of embryo width (mean ± S.D., $N = 7$ embryos, $p = 9.8 \times 10^{-7}$) compared to $22.4 \pm 9.8\%$ ($N = 29$) as expected (11). Under that condition, the position, at which the oscillations started, was not strongly nor significantly altered (Figure S2C). Overall, these three results support our model: microtubule dynamics sense centrosome position and the extent of the active region. By this mean, this latter controls the position at which oscillation onset starts.

**The position, at which pulling forces burst, is resistant to variations in embryo length.**

Because the astral microtubules sense the distance from the centrosomes to the cortex, the proposed mechanism may also make the positional regulation of the cortical pulling forces dependent on the cell geometry. To this respect, it was reminiscent of the yeast positional check point (47) and of adaptation of cell division plane positioning to variations in cell shape observed in other contexts and organisms (41, 48, 49). Our expanded model suggested that the posterior centrosome's position at oscillation onset would only weakly depend upon the length of the embryo (Figure 4A). To investigate experimentally this prediction, we depleted C27D9.1 and CID-1 by *RNAi* to obtain longer and shorter embryos (Figure 4D), respectively. In both cases, the embryos were viable and showed no other phenotype. We measured the variations in the timing and positioning of oscillation onset with respect to the variations in embryo length and we normalized the shifts using control averages as references to enable comparisons (Methods). For both cases, we fitted a linear regression, measuring oscillation onset timing slope about 7 times larger than that of the oscillation onset position (Figure 4BC). It was indicative that the positional switch at about 70% of AP axis was conserved despite variations in embryo length. The very low slope of the linear regression in Figure 4C suggested that the

positional control could make the position at which cortical force bursts resistant to changes in embryo length. Interestingly, Farhadifar et al. showed that both the final spindle length and the division plane position scale linearly with embryo length within different *C. elegans* natural isolates and beyond within several nematode species (50). These correlations may be either enforced by natural selection or due to some robustness mechanisms, two non-exclusive possibilities. The cortical pulling forces play an important role in these both positionings and the positional regulation exhibited here may be one of the mechanisms accounting for such a robustness. In contrast, the timing of oscillation onset showed a positive correlation with embryo length, with a larger linear regression slope than the one measured for the position. This is consistent with longer embryos requesting a larger duration of posterior displacement to reach the critical 70% of AP axis position for oscillation onset (13). Overall, it also supports the idea that the positioning and timing of oscillation initiation are not linked with each other at anaphase onset.

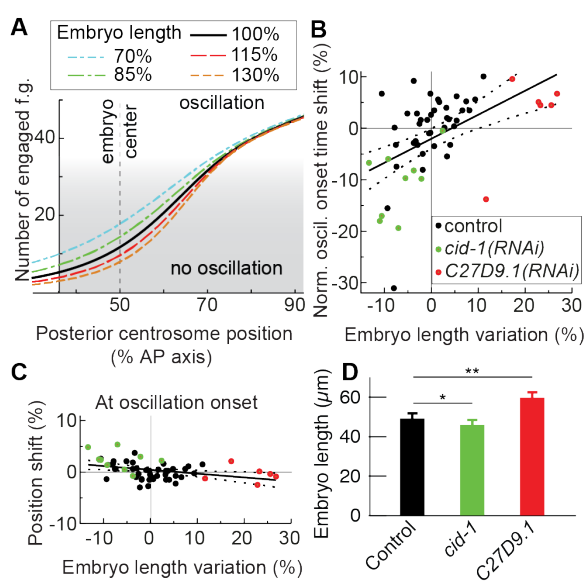

**Figure 4: Embryo length has less effect on oscillation onset position than on its timing.**
(A) Modelled number of engaged force generators versus the posterior displacement of the centrosome along the AP axis as a percentage of embryo length. The line colours indicate the embryo length: untreated embryos are black; the shorter embryos corresponding to the one produced by *cid-1(RNAi)* are shown in blue and green; and the longer ones from *c27d9.1(RNAi)* are shown in red and orange. The parameters used are listed in Table S5. Grey shading indicates when the number of engaged force generators was too low to permit oscillation. (B-C) Variations in embryo lengths as compared to the control (normalized by the average length, see Methods) are shown here (B) versus the shift in oscillation onset timing normalized by the control's average pro-metaphase and metaphase duration; and (C) versus the shift in the posterior centrosome's position at oscillation onset as compared to the control. The solid black lines indicate the linear least square fits, with slopes of 0.47 ± 0.11 ($p = 5 \times 10^{-5}$ compared to null slope), -0.07 ± 0.02 ($p = 0.005$), respectively. The black dashed lines are the standard errors of the mean. Dots indicate individual embryos, and the average control values (0 shift) are thin black lines. (D) Embryo lengths in control and embryos treated by RNAi to vary their lengths. Error bars indicate SD, and stars indicate significant differences (Methods). We measured $N = 9$ *cid-1(RNAi)*, $N = 6$ *c27d9.1(RNAi)*, and $N = 39$ control embryos with GFP::γ-tubulin-labelled centrosomes at 23ºC. The control embryos used are the same as shown in Figure 3.

**Centrosome position and time/mitotic progression both contribute to oscillation and broadly pulling force regulation.**

We found that the oscillation onset is controlled by the position of the posterior centrosome while its die-down timing is dependent on the time control (mitotic progression). Beyond these dominant effects, we however noticed that the oscillation might start while the centrosome was close to 70%, rather than when it reached this position (Table 1). We suspected that a combination of both temporal and positional controls was at work. We set to recapitulate this regulation by completing our model and validating it. We included the temporal control within the expanded model, as previously done in the initial model, i.e. through the decrease of microtubule force-generator detachment rate (off-rate), which is the inverse of processivity (16). In further details, we made the microtubule-force generator equilibrium–constant dependent on time through the decreasing off-rate. We will from now refer to this model as the "full-expanded model" (Suppl. Model). In contrast to the initial model, the force generator on-rate is not constant and depends on the number of microtubules available at the cortex for binding a force generator and on the extent of the active region. It also depends on the on-rate imbalance reflecting polarity (14). Using the full-expanded model, we computed the stability diagram corresponding to centrosome oscillation, which depended on both temporal and positional control parameters: the force generator processivity and the position of the posterior centrosome, respectively. Non-stable region (Figure 5A, blue region) corresponded to parameter sets in which oscillation developed. In contrast, in stable ones (Figure 5A, white regions), the system was overdamped, making oscillation die down. It suggested that to enable oscillation, in addition to a large enough force generator processivity (16), the posterior centrosome needed to be close to the posterior tip of the embryo, which supported our positional switch hypothesis investigated above (Figure 5A, thick blue curve). This led to a dual control of the pulling forces. While this model offered a clear view about oscillation onset, we could only infer predictions about its die-down from the return-to-stability curve, which corresponded to when the system dampens out oscillation (Figure 5A, green curve). This was because no detailed model relates mitotic progression/time and processivity. Interestingly, the steeper slope of this return-to-stability curve compared to the oscillation onset curve suggested nevertheless that the posterior centrosome's position was likely to more strongly influence oscillation onset than die-down (Figure 5A). This was consistent with the equal anaphase onset to oscillation die-down duration observed in *such-1(h1960)* mutant and its control (Table 1).

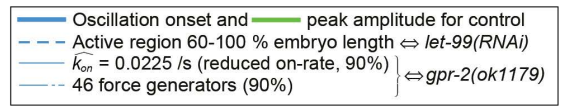
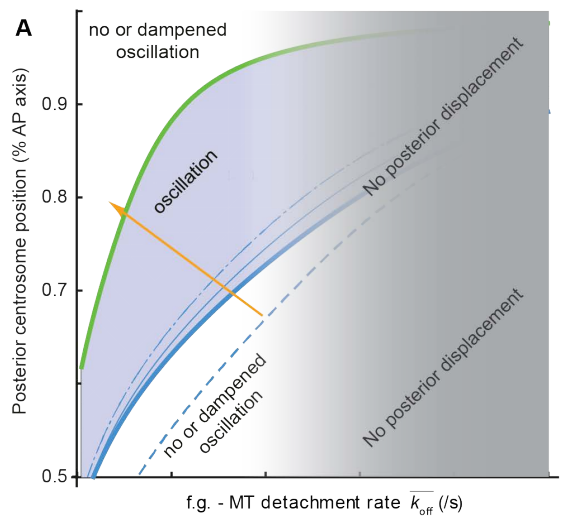
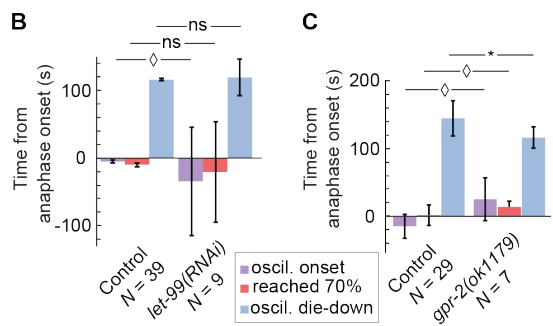

**Figure 5: Two independent controls both contribute to oscillations.**
(A) Stability diagram of the full-expanded model as a function of the detachment rate (off-rate $\overline{k_{off}}$, inverse of the processivity, *x*-axis) and of the position of the centrosome as a percentage of embryo length (*y*-axis). The unstable region (blue) corresponds to the values of off-rate and posterior-centrosomal position enabling oscillation development. The critical values are marked by the thick blue and green lines for parameters corresponding to control condition. Thin lines with various dashing correspond to oscillation onset's critical curves, for varied parameters, indicated by value and % of control condition, and corresponding to experimental perturbations. The orange arrow indicates the typical "phase" trajectory during mitosis, based on the parameters used in this study. The greyed-out area depicts that above a detachment rate threshold, the posterior displacement of the spindle/posterior centrosome no longer occurs (orange curve in Figure 7C). The centrosome needs to reach a position that is posterior enough to enable oscillations while force generators must display a high enough processivity (measured to 1-2 s$^{-1}$ in metaphase (14)). The parameters used are listed in Table S5. (B-C) Timings of oscillation onset, die-down, and posterior centrosome arrival at 70% of embryo length (B) when the size of the active region is changed in *let-99(RNAi)* (*N* = 9) compared to untreated (*N* = 39) embryos at 23°C, and (C) upon depletion of *active* force generators (f.g.) in *gpr-2(ok1179)* mutants (*N* = 7) compared to *N* = 29 untreated embryos at 18°C. All embryos display GFP::γ-tubulin labelling of the centrosomes. Error bars indicate SD and diamonds or stars indicate significant differences (Methods).

To challenge this full-expanded model, we again investigated the role of players in each control (Figure 3A). We first observed that an anterior shift of the boundary position of the active region

17

by *let-99(RNAi)* did not alter the oscillation die-down timing significantly with respect to the control, while oscillation onset happened earlier (Figure 5B) and when the posterior centrosome was positioned more anteriorly (Figure 3C). This was consistent with model predictions (Figure 5A, blue dashed line) and expected since oscillation die-down depended on time control (Figure 3A). Secondly and in contrast, we decreased the number of active force generators at the cortex, using *gpr-2(ok1179)* mutant, and measured a precocious oscillation die-down (Figure 5C), while oscillation onset position was unchanged (Figure S2C). This result was predicted by the model, disregarding whether we modelled the decrease in active force generators through their number (11, 15) or their on-rate binding to microtubules (14) (Figure 5A, chain and continuous thin blue lines, respectively). Overall, the full-expanded model correctly accounted for phenotypes resulting from perturbations affecting a player either related to positional module or to time module (Figure 3A). Interestingly, the resulting phenotype was affecting mostly, either the posterior centrosome's position at oscillation onset or the timing of oscillation die-down, but not both, suggesting some independence of the two pathways. It is likely caused by the different players involved in each control. This dual regulation calls for re-examining how the final position of the spindle is set, superseding the initial model, in which we proposed that only the force generator number and dynamics contributed (16).

**LET-99 spatial restriction of active force generators sets the spindle's final position.**

Having ascertained the full-expanded model using the oscillation onset, we used it to investigate the mechanism connecting the cortical polarity, through its downstream effector LET-99, to the final centrosomes' position. Indeed, both spindle poles' and centre's final positions set the position of the cytokinesis furrow (8). They depend upon cortical pulling forces and therefore the positional and temporal controls over these forces were expected to make the spindle's final position robust to some extent. In contrast, in our initial model, we suggested that the final posterior centrosome's position resulted from a balance between the cortical pulling forces and centring forces, the latter being modelled by a spring (16, 33). To investigate spindle's final positioning, we simulated the posterior displacement using our full-expanded model (Methods). We could reproduce the global kinematics of posterior displacement: slow prior to anaphase then accelerating afterward (Figure S3E) as we previously observed experimentally (33).

LET-99 sets the boundary of the active region and was proposed to set spindle's final position downstream of polarity (19, 20). We foresaw that the full-expanded model proposed here could recapitulate the mechanism of this control. To test this, we simulated the posterior centrosome's displacement during mitosis with different boundary positions of the active region and observed that the final position of the posterior centrosome was displaced anteriorly when the boundary of the posterior active crescent moved anteriorly, so long as the region was large enough to initiate posterior displacement in the first place (Figure 6C, solid lines). This result was confirmed experimentally in *let-99(RNAi)*-treated embryos (Figure 6A), and is consistent with previous observations (20). The prediction and experimental observations differed from the initial model ones, which stated that under similar cortical forces, the posterior displacement would be the same disregarding the cortical distribution of the force generators (Figure 6C, dashed line).

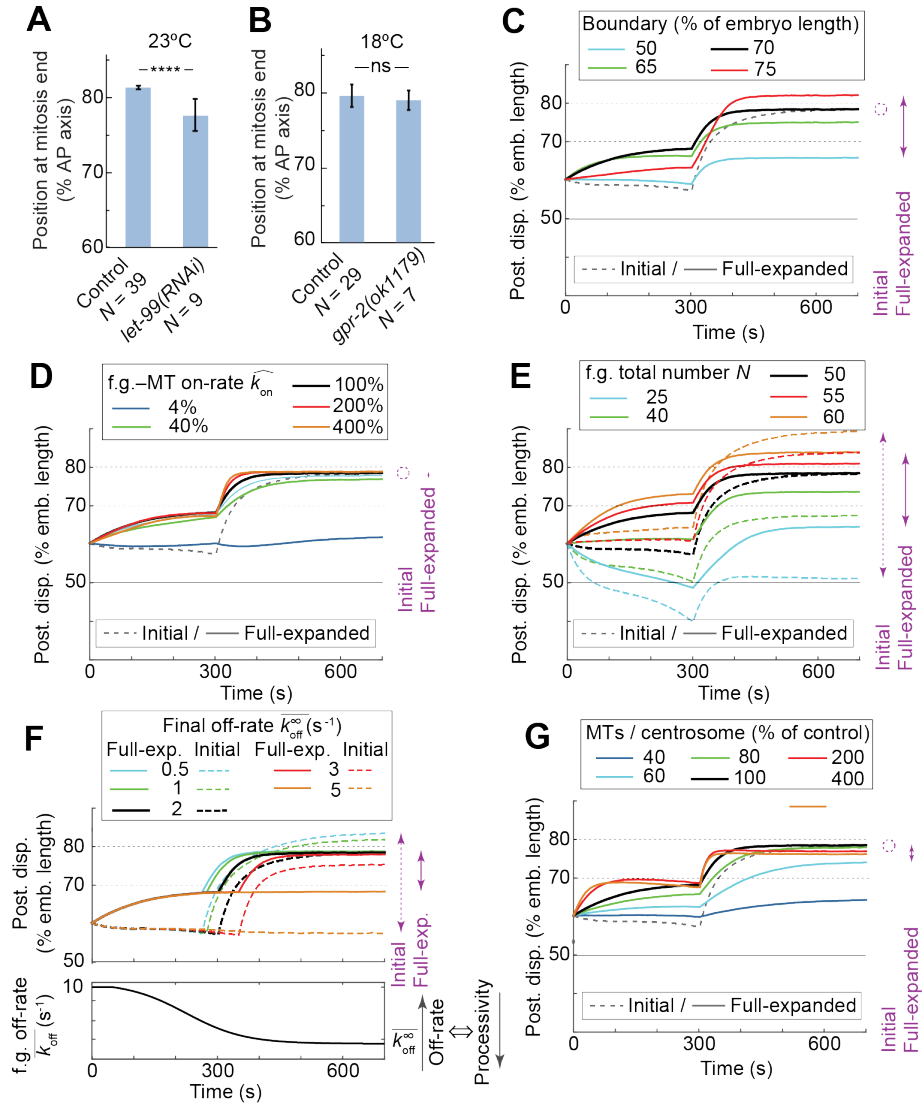

**Figure 6: Active region boundary position sets the spindle's final position.**
(A-B) Posterior centrosome position at oscillation die-down (A) upon anteriorly extending the active region by *let-99(RNAi)* ($N$ = 9 embryos) compared to control ($N$ = 39 embryos) at 23°C, and (B) upon decreasing cortical force generation using mutant *gpr-2(ok1179)* ($N$ = 7) compared to $N$ = 29 control embryos at 18°C. In all cases, centrosomes were labelled by GFP::γ-tubulin. Error bars indicate SD, and stars indicate significant differences (Methods). (C-G) Posterior displacement of the posterior centrosome averaged over 25 simulation runs with respectively varied: (C) the position of the boundary of the active region; (D) the binding rate (on-rate) of the force generators and microtubules $\widehat{k_{on}}$, whose asymmetry may encode the polarity (14); (E) the total number of active force generators available at the posterior active region $N$ (active, i.e. currently pulling or ready to do so when meeting a microtubule); (F) the force generator's (f.g.) final detachment rate (off-rate, the inverse of the processivity) $\overline{k_{off}^{\infty}}$; and (G) the number of microtubules emanating from each centrosome (in % of control, $M$ = 3000). When it does not depend on the parameter considered, the initial model is shown by a grey dashed line. In all cases, the control values are black; lower values are blue and green; and the higher values are red and orange. The dispersions of the final values for each case are represented on the right side of the plots by purple arrows in dashed or solid lines according to the model used, and a larger span in the plot reveals a lack of robustness to parameter variations. A circle is used when the parameter has no effect on the final value.

LET-99 may act either by concentrating force generators in the posterior active region or by spatially restricting them to a posterior crescent. To decide between these mechanisms, we tested how sensitive was the centrosome's final position with respect to the precise final count or dynamics of the force generators. Our simulation suggested a reduced dependency, in contrast with the initial model, provided that a threshold force was generated to enable posterior displacement. In particular, we investigated the impact of the total number of force generators available (Figure 6E), of their final detachment rate (off-rate) that reflected the time/mitotic progression control (Figure 6F) and of their binding rate to microtubules (on-rate), recently suggested to reflect polarity (14) (Figure 6D). In this latter case, we scaled up or down the on-rate in similar ways both on anterior and posterior sides. Experimentally, we observed that *gpr-2(ok1179)*, a mutant that partially reduced the cortical forces (11, 15), caused no significant shift in spindle's final position (Figure 6B). This full-expanded model's robustness was attributed to a smaller increase in cortical pulling forces after crossing the boundary position of the active region (Figure S3D) since the increasing processivity was tempered by a saturation of microtubule contact count. Along the same line, it was likely that the force generator's final processivity was disturbed in *such-1(h1960)*, the mutant with delayed anaphase onset (22) and as predicted, the final position of the posterior centrosome was conserved (Table 1).

Microtubule is a key player of our positional mechanism and is proposed to contribute to anaphase oscillation through a modulation of their cortical contacts (23). We thus wondered how important was the total number of microtubules in setting the spindle's final position. We simulated a scaling down of their number and observed that, above a threshold needed for posterior displacement, their precise number was again unimportant (Figure 6G). Consistently, we found no strong alteration of posterior centrosome's final position upon partially depleting SPD-2 when compared to the shift measured in its position at oscillation onset (Table 1). Interestingly, we also observed an optimum value to maximize displacement (Figure 6G, red and green curves below the black one). We overall conclude that the proposed mechanism in which the distribution of force generators is spatially restricted to the posterior crescent perfectly accounts for the connection between the LET-99 band and the final position of the posterior centrosome and consequently of the spindle (Figure 3A, yellow). Furthermore, it also suggests that this position is resistant to modest changes in the parameters related to microtubules or force generators, as long as they reach threshold values needed for posterior displacement.

## DISCUSSION

We measured the spatial distribution of microtubule contacts at the cell cortex and found it uneven, with higher concentration in the regions closer to the centrosomes. As a direct consequence and in combination with the restriction of cortical pulling force generation to a posterior crescent (20), the number of microtubule available to cortical force generators displayed a strong increase linked to the spindle's posterior displacement, with a switch-like behaviour. This in turn made the cortical pulling forces dependent on the position of the centrosomes. Astral microtubule dynamics create a *positional control* over forces, especially over the burst at anaphase onset and the consequent spindle oscillations, in addition to the previously described regulation by pulling force generator dynamics (*temporal control*). We quantitatively recapitulated this novel regulation into an expanding of our initial spindle

oscillation and posterior displacement model (16). This positional regulation superimposes firstly to the force imbalance, reflecting polarity, essential for the spindle to be displaced posteriorly (11, 15, 51) and secondly to the temporal control related to mitotic progression (22), which sets the velocity of displacement and consequently its duration. To this respect, in *C. briggsae*, we observed lower amplitude oscillation and a 30 s delay in posterior centrosome reaching 70% of embryo length (13).

Importantly, this dual regulation of the cortical pulling forces perfectly accounts for the robustness of the spindle's final positioning. We prove that restricting spatially force generation to a posterior crescent is key, while the precise level of active force generators or microtubule contacts at the cortex is of lower importance. By precise level, we mean that minimal numbers of microtubules and active force generators are required to ensure posterior displacement, but beyond that constraint, the values do not need to be finely tuned and modest variations can be buffered. Understanding robustness by accounting for a spatial distribution and regulation of force generation is, in its principle, similar to the contribution of both the asymmetric microtubule array and a LET-99 band to force regulation during centration and orientation of the pronuclei-centrosome complex (52). Indeed, we propose here that the mechanism linking the LET-99 domain and spindle's final position relies on spatial restriction of active force generators to a posterior crescent. In contrast and because the spindle's final position is robust to modest active force generator variations, it is unlikely that LET-99 acts by concentrating the force generators in posterior crescent. The imbalance in force generator number, reflecting GPR-1/2 posterior concentration (12, 13), is likely unrelated to LET-99 band. Consistently, an extra shift of spindle's final positioning towards anterior was observed previously upon treating a *let-99(or81)* mutant with *gpr-1/2(RNAi)* (53).

The positional control of cortical pulling forces contributes also to ensuring the correct positioning of cytokinesis cleavage furrow, essential for the correct distribution of cell polarity cues and thus daughter cell fates (1, 2, 5). Indeed, in late mitosis, its position is signalled by two parallel pathways: firstly, by the spindle poles via astral microtubules and secondly by the central spindle (54). For the latter, the signalling comes from the central spindle's position when the spindle has reached its full elongation, which is controlled by the pulling forces (55). Overall, by controlling the position of the posterior centrosome and where pulling force burst causing elongation happens, the proposed mechanism, based on microtubule network sensing cell geometry, contributes to offering robustness to the cleavage furrow positioning.

Quantitative genetic studies in *C. elegans* have found an about threefold lower persistence of new mutations on the phenotypic trait positioning the division plane compared to the trait displaying the time between first oscillation peak and mid-elongation (50). We here measured the related parameter, duration of oscillations. As explained above, peak amplitude corresponds to the return–to–stability in our diagram (Figure 5A) and behaves similarly to oscillation die-down. We found a spindle's final position with a reduced dependency on variations in force generator number and dynamics as reported by e.g. oscillation duration. This is consistent that the proposed robustness mechanism de-correlates the spindle's final position, essential likely to viability, from timings that appear less evolutionary constrained. Along the same line, we recently performed a comparative study between two nematode cousins (*C. elegans* and *C.*

*briggsae*) (13). Because of a duplication of the regulator of force generators in *C. elegans*, GPR-1 and GPR-2, while *C. briggsae* only displays GPR-2, the pulling force regulation is altered leading to a delay in posterior displacement in the latter species. The posterior centrosome reaching of 70% of AP axis is delayed by 30 s and so does the oscillation onset, both with respect to anaphase onset. In contrast, the oscillation die-down timing is correlated with the one of anaphase onset in both species. Our dual positional and time control perfectly accounts for the buffering of timing differences in centrosome kinematics, to ensure robust oscillation onset position and similar spindle's final positioning in both species. That evolution of the essential *gpr* gene was likely made possible by the positional control and the resulting resistance to variations in force generator quantity and dynamics. Interestingly, cross-species insertion of *gpr* genes modulates oscillation amplitude but preserves the positional control, which is consistent with our *gpr-2(ok1179)* experiment (13). The robustness of spindle's final positioning is likely to be true in more than just these two species.

Finally, the observed positional switch is caused by astral microtubules, and more precisely by the number of microtubule cortical contacts, which reflects the distance between the centrosome and the cortex. Indeed, said distance is measured in "units of microtubule dynamics". Astral microtubules provide feedback about the posterior centrosome position to the cortical pulling forces which suggests the existence of a mechano-sensing pathway. Such a property already enables classic mechanisms for creating centring (40, 56) or other shape-dependent mechanisms (41, 57, 58). However, such mechanisms were mostly inferred from cell-level measurements. In contrast, here, the distribution of the microtubule-end contacts located at the cortex was obtained from microscopic measurements. We observed a density ratio of about 2 between the regions with the most and least microtubule contacts at a given time, and this ratio represents the sensitivity to centrosomal position. From a theoretical point of view, considering the ellipsoidal shape of the *C. elegans* embryo and the microtubule dynamics measurements performed elsewhere, the predicted maximal ratio is 1.64 (Suppl. Model). Our experimental result is close to this prediction, suggesting that the microtubule dynamics parameters are optimal for the positional control that we discussed here.

## CONCLUSION

This study of pulling force regulation by the spindle position was grounded on studying experimentally and by modelling the microtubules at microscopic– and cell–scale, supplemented by investigating the precise timing of transverse oscillation onset with respect to the anaphase and the posterior centrosome positioning at that moment in the *C. elegans* embryo. It has highlighted the key role of microtubule dynamics in probing the boundary of the active force generator region. In particular, microtubules create a *positional control* over the spindle oscillation, which acts in addition to the previously described regulation by pulling force generator dynamics (*temporal control*). Both controls contribute independently to the switch to prevent premature pulling force burst. The finding of this supplementary positional control paves the way to a novel understanding of the mitosis choreography mechanism, supplementing the regulation by the only cell cycle.

In particular, our proposed positional control enables to connect the spindle's final position to the cortical polarity cues. Indeed, LET-99 restricts spatially the force generators to a posterior

crescent, and in turn, microtubule dynamics read out the boundary of this region to properly position the centrosomes and thus the spindle at the end of the mitosis. Interestingly, It also makes this position resistant to modest variations in microtubule quantity or in force generator dynamics or number, provided that these are above the required thresholds. This guarantees a correct cytokinesis furrow positioning and consequently a proper polarity cue distribution and daughter cell fates (3, 4). Such a robustness buffered changes in cortical pulling force levels and timings between the *C. elegans* and *C. briggsae* nematodes. This permitted substantial modifications in the essential *gpr-1/2$^{LGN}$* genes, whose proteins are part of the complex that generates cortical pulling forces (13). Finally, the observed positional switch is caused by astral microtubules. They provide feedback about the posterior centrosome position with respect to the posterior crescent, to the cortical pulling forces. It suggests the existence of a mechano-sensing pathway. This finding is a novel example of a microfilament-based system that controls essential aspects of cell division. In contrast with robustness resulting from classic biochemical signalling pathways (59, 60), this mechanism is based solely on cell mechanics and component dynamics.

**Author contributions**

Conceptualization, JP, HB, and MD; Methodology, HB, LC, and JP; Software, JP and HB; Validation / Formal Analysis, HB, SP, and LC; Investigations / Data Curation, HB, SP, LC, and JP; Writing – Original Draft, HB and JP; Writing - Review & Editing, HB, LC, MD, and JP; Supervision, JP.

**Supplemental information**

Figure S1: Validation of the strain and of the analysis pipeline used to measure microtubule contact density at the cell cortex.

Figure S2: Microtubule number affects the oscillation onset position, while force generator quantity above a threshold does not.

Figure S3: Simulation of the displacement of the posterior centrosome using full-expanded model.

Figure S4: Landmarks used to set reference times and oscillation characteristics.

Table S5: Parameters used for modelling and simulations.

Supplementary model

Supplementary method: The "landing" assay

**References**

1. Betschinger, J., and J. A. Knoblich. 2004. Dare to be different: asymmetric cell division in Drosophila, C. elegans and vertebrates. Curr Biol 14:R674-685.
2. Gillies, T. E., and C. Cabernard. 2011. Cell division orientation in animals. Curr Biol 21:R599-609.
3. Gonczy, P. 2008. Mechanisms of asymmetric cell division: flies and worms pave the way. Nat Rev Mol Cell Biol 9:355-366.
4. Morin, X., and Y. Bellaiche. 2011. Mitotic spindle orientation in asymmetric and symmetric cell divisions during animal development. Dev Cell 21:102-119.
5. White, E. A., and M. Glotzer. 2012. Centralspindlin: At the heart of cytokinesis. Cytoskeleton (Hoboken) 69:882-892.
6. Rappaport, R. 1971. Cytokinesis in animal cells. Int Rev Cytol 31:169-213.
7. Knoblich, J. A. 2010. Asymmetric cell division: recent developments and their implications for tumour biology. Nat Rev Mol Cell Biol 11:849-860.
8. Bringmann, H., and A. A. Hyman. 2005. A cytokinesis furrow is positioned by two consecutive signals. Nature 436:731-734.
9. McNally, F. J. 2013. Mechanisms of spindle positioning. Journal of Cell Biology 200:131-140.
10. Nguyen-Ngoc, T., K. Afshar, and P. Gonczy. 2007. Coupling of cortical dynein and G alpha proteins mediates spindle positioning in Caenorhabditis elegans. Nat Cell Biol 9:1294-1302.
11. Colombo, K., S. W. Grill, R. J. Kimple, F. S. Willard, D. P. Siderovski, and P. Gonczy. 2003. Translation of polarity cues into asymmetric spindle positioning in Caenorhabditis elegans embryos. Science 300:1957-1961.
12. Park, D. H., and L. S. Rose. 2008. Dynamic localization of LIN-5 and GPR-1/2 to cortical force generation domains during spindle positioning. Dev Biol 315:42-54.
13. Riche, S., M. Zouak, F. Argoul, A. Arneodo, J. Pecreaux, and M. Delattre. 2013. Evolutionary comparisons reveal a positional switch for spindle pole oscillations in Caenorhabditis embryos. J Cell Biol 201:653-662.
14. Rodriguez Garcia, R., L. Chesneau, S. Pastezeur, J. Roul, M. Tramier, and J. Pecreaux. 2017. Dynein dynamics at the microtubule plus-ends and cortex during division in the C. elegans zygote. bioRxiv:118604.
15. Grill, S. W., J. Howard, E. Schaffer, E. H. Stelzer, and A. A. Hyman. 2003. The distribution of active force generators controls mitotic spindle position. Science 301:518-521.
16. Pecreaux, J., J. C. Roper, K. Kruse, F. Julicher, A. A. Hyman, S. W. Grill, and J. Howard. 2006. Spindle oscillations during asymmetric cell division require a threshold number of active cortical force generators. Curr Biol 16:2111-2122.
17. Grill, S. W., K. Kruse, and F. Julicher. 2005. Theory of mitotic spindle oscillations. Phys Rev Lett 94:108104.

18. Tsou, M. F., A. Hayashi, L. R. DeBella, G. McGrath, and L. S. Rose. 2002. LET-99 determines spindle position and is asymmetrically enriched in response to PAR polarity cues in C. elegans embryos. Development 129:4469-4481.
19. Wu, J. C., E. B. Espiritu, and L. S. Rose. 2016. The 14-3-3 protein PAR-5 regulates the asymmetric localization of the LET-99 spindle positioning protein. Dev Biol.
20. Krueger, L. E., J. C. Wu, M. F. Tsou, and L. S. Rose. 2010. LET-99 inhibits lateral posterior pulling forces during asymmetric spindle elongation in C. elegans embryos. J Cell Biol 189:481-495.
21. Howard, J. 2006. Elastic and damping forces generated by confined arrays of dynamic microtubules. Phys Biol 3:54-66.
22. McCarthy Campbell, E. K., A. D. Werts, and B. Goldstein. 2009. A cell cycle timer for asymmetric spindle positioning. PLoS Biol 7:e1000088.
23. Kozlowski, C., M. Srayko, and F. Nedelec. 2007. Cortical microtubule contacts position the spindle in C. elegans embryos. Cell 129:499-510.
24. Espiritu, E. B., L. E. Krueger, A. Ye, and L. S. Rose. 2012. CLASPs function redundantly to regulate astral microtubules in the< i> C. elegans</i> embryo. Developmental biology.
25. Gönczy, P., J. M. Bellanger, M. Kirkham, A. Pozniakowski, K. Baumer, J. B. Phillips, and A. A. Hyman. 2001. zyg-8, a gene required for spindle positioning in C. elegans, encodes a doublecortin-related kinase that promotes microtubule assembly. Developmental cell 1:363.
26. Brenner, S. 1974. The genetics of Caenorhabditis elegans. Genetics 77:71-94.
27. Srayko, M., A. Kaya, J. Stamford, and A. A. Hyman. 2005. Identification and characterization of factors required for microtubule growth and nucleation in the early C. elegans embryo. Dev Cell 9:223-236.
28. Oegema, K., A. Desai, S. Rybina, M. Kirkham, and A. A. Hyman. 2001. Functional analysis of kinetochore assembly in Caenorhabditis elegans. The Journal of Cell Biology 153:1209-1226.
29. Tarailo, M., R. Kitagawa, and A. M. Rose. 2007. Suppressors of spindle checkpoint defect (such) mutants identify new mdf-1/MAD1 interactors in Caenorhabditis elegans. Genetics 175:1665-1679.
30. Riddle, D. L. 1997. C. elegans II. Cold Spring Harbor Laboratory Press, Plainview, N.Y.
31. Tinevez, J. Y., J. Dragavon, L. Baba-Aissa, P. Roux, E. Perret, A. Canivet, V. Galy, and S. Shorte. 2012. A quantitative method for measuring phototoxicity of a live cell imaging microscope. Methods in Enzymology 506:291-309.
32. Li, S., S. Besson, C. Blackburn, M. Carroll, R. K. Ferguson, H. Flynn, K. Gillen, R. Leigh, D. Lindner, M. Linkert, W. J. Moore, B. Ramalingam, E. Rozbicki, G. Rustici, A. Tarkowska, P. Walczysko, E. Williams, C. Allan, J. M. Burel, J. Moore, and J. R. Swedlow. 2016. Metadata management for high content screening in OMERO. Methods 96:27-32.
33. Pecreaux, J., S. Redemann, Z. Alayan, B. Mercat, S. Pastezeur, C. Garzon-Coral, A. A. Hyman, and J. Howard. 2016. The Mitotic Spindle in the One-Cell C. elegans Embryo Is Positioned with High Precision and Stability. Biophys J 111:1773-1784.
34. Roul, J., J. Pecreaux, and M. Tramier. 2015. Method for controlling a plurality of functional modules including a multi-wavelength imaging device, and corresponding control system. U. D. R. Cnrs - Centre National De La Recherche Scientifique, editor.
35. Kalman, R. E. 1960. A New Approach to Linear Filtering and Prediction Problems. Journal of Basic Engineering 82:35-45.

36. Jaqaman, K., D. Loerke, M. Mettlen, H. Kuwata, S. Grinstein, S. L. Schmid, and G. Danuser. 2008. Robust single-particle tracking in live-cell time-lapse sequences. Nat Methods 5:695-702.
37. Pecreaux, J., C. Zimmer, and J. C. Olivo-Marin. 2006. Biophysical active contours for cell tracking I: Tension and bending. In 2006 Ieee International Conference on Image Processing, Icip 2006, Proceedings, October 8-11 2006, Atlanta, GA, USA. 1949-1952.
38. Hosea, M. E., and L. F. Shampine. 1996. Analysis and implementation of TR-BDF2. Applied Numerical Mathematics 20:21-37.
39. Dogterom, M., and B. Yurke. 1998. Microtubule dynamics and the positioning of microtubule organizing centers. Physical Review Letters 81:485-488.
40. Dogterom, M., J. W. Kerssemakers, G. Romet-Lemonne, and M. E. Janson. 2005. Force generation by dynamic microtubules. Curr Opin Cell Biol 17:67-74.
41. Minc, N., D. Burgess, and F. Chang. 2011. Influence of cell geometry on division-plane positioning. Cell 144:414-426.
42. Mitchison, T., and M. Kirschner. 1984. Dynamic instability of microtubule growth. Nature 312:237-242.
43. O'Rourke, S. M., S. N. Christensen, and B. Bowerman. 2010. Caenorhabditis elegans EFA-6 limits microtubule growth at the cell cortex. Nat Cell Biol 12:1235-1241.
44. O'Toole, E. T., K. L. McDonald, J. Mantler, J. R. McIntosh, A. A. Hyman, and T. Muller-Reichert. 2003. Morphologically distinct microtubule ends in the mitotic centrosome of Caenorhabditis elegans. J Cell Biol 163:451-456.
45. Labbe, J. C., E. K. McCarthy, and B. Goldstein. 2004. The forces that position a mitotic spindle asymmetrically are tethered until after the time of spindle assembly. J Cell Biol 167:245-256.
46. Koonce, M. P., and I. Tikhonenko. 2012. Dynein Motor Mechanisms. In Dyneins : structure, biology and disease. S. M. King, editor. Academic Press, Amsterdam ; Boston. xv, 639 p.
47. Moore, J. K., V. Magidson, A. Khodjakov, and J. A. Cooper. 2009. The spindle position checkpoint requires positional feedback from cytoplasmic microtubules. Current Biology 19:2026-2030.
48. Tanimoto, H., A. Kimura, and N. Minc. 2016. Shape-motion relationships of centering microtubule asters. J Cell Biol 212:777-787.
49. Thery, M., A. Jimenez-Dalmaroni, V. Racine, M. Bornens, and F. Julicher. 2007. Experimental and theoretical study of mitotic spindle orientation. Nature 447:493-496.
50. Farhadifar, R., J. M. Ponciano, E. C. Andersen, D. J. Needleman, and C. F. Baer. 2016. Mutation Is a Sufficient and Robust Predictor of Genetic Variation for Mitotic Spindle Traits in Caenorhabditis elegans. Genetics 203:1859-1870.
51. Grill, S. W., P. Gonczy, E. H. Stelzer, and A. A. Hyman. 2001. Polarity controls forces governing asymmetric spindle positioning in the Caenorhabditis elegans embryo. Nature 409:630-633.
52. Coffman, V. C., M. B. McDermott, B. Shtylla, and A. T. Dawes. 2016. Stronger net posterior cortical forces and asymmetric microtubule arrays produce simultaneous centration and rotation of the pronuclear complex in the early Caenorhabditis elegans embryo. Mol Biol Cell 27:3550-3562.
53. Tsou, M. F., A. Hayashi, and L. S. Rose. 2003. LET-99 opposes Galpha/GPR signaling to generate asymmetry for spindle positioning in response to PAR and MES-1/SRC-1 signaling. Development 130:5717-5730.

54. Bringmann, H., C. R. Cowan, J. Kong, and A. A. Hyman. 2007. LET-99, GOA-1/GPA-16, and GPR-1/2 are required for aster-positioned cytokinesis. Curr Biol 17:185-191.
55. Lewellyn, L., J. Dumont, A. Desai, and K. Oegema. 2010. Analyzing the effects of delaying aster separation on furrow formation during cytokinesis in the Caenorhabditis elegans embryo. Mol Biol Cell 21:50-62.
56. Wuhr, M., S. Dumont, A. C. Groen, D. J. Needleman, and T. J. Mitchison. 2009. How does a millimeter-sized cell find its center? Cell Cycle 8:1115-1121.
57. Daga, R. R., and P. Nurse. 2008. Interphase microtubule bundles use global cell shape to guide spindle alignment in fission yeast. J Cell Sci 121:1973-1980.
58. Thery, M. 2010. Micropatterning as a tool to decipher cell morphogenesis and functions. J Cell Sci 123:4201-4213.
59. McIntosh, J. R., M. I. Molodtsov, and F. I. Ataullakhanov. 2012. Biophysics of mitosis. Quarterly reviews of biophysics 45:147-207.
60. Musacchio, A. 2011. Spindle assembly checkpoint: the third decade. Philosophical transactions of the Royal Society of London. Series B, Biological sciences 366:3595-3604.

# Supplementary Figures

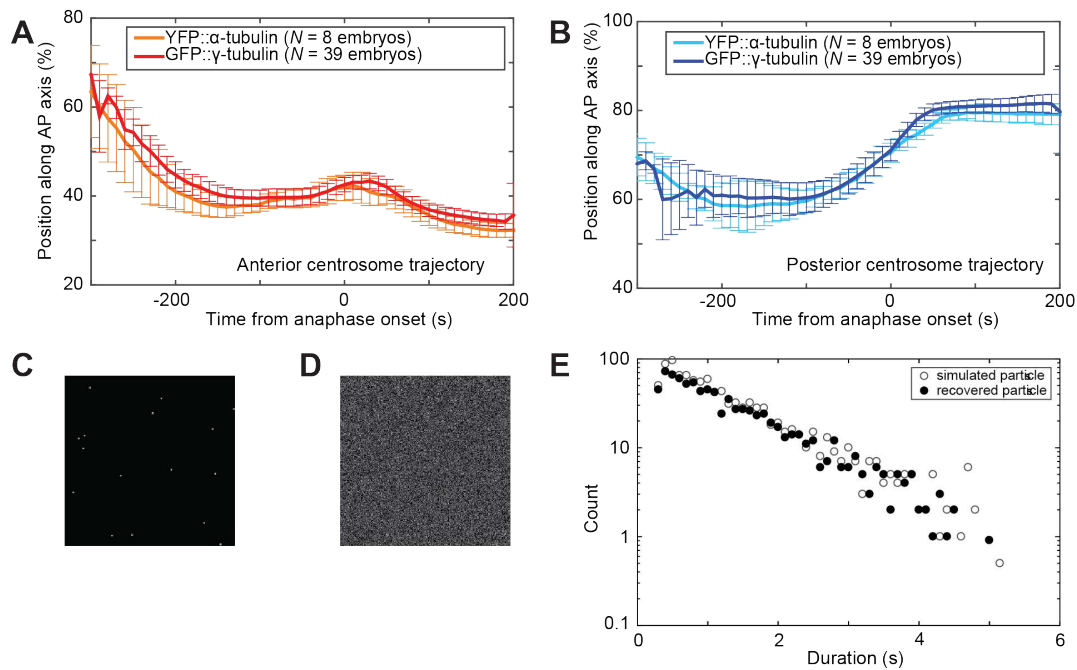

**Figure S1: Validation of the strain and of the analysis pipeline used to measure microtubule contact density at the cell cortex.**
(A-B) Comparison of the (A) anterior and (B) posterior centrosome trajectories during mitosis between two *C. elegans* strains at 23°C, one with YFP::α-tubulin–labelled microtubules (orange and light-blue lines, $N$ = 8 embryos) and the other with GFP::γ-tubulin-labelled centrosomes (red and dark-blue lines, $N$ = 39 embryos). The positions along the AP axis are shown as a percentage of embryo length. (C-D) Fabricated images with particles of known dynamics and their analysis to validate the image-processing pipeline (Suppl. Methods). (C) We simulated particles (bright spots) that mimic the microtubule contacts at the cortex. (D) We then added noise to mimic the background observed experimentally. (E) Finally, we compared the histograms of particle lifetime between values provided to the simulation (open circles) and the ones recovered through our analysis (closed symbols). The parameters used to fabricate *in silico* microtubule cortical contact dynamics images are listed in Supplementary Methods.

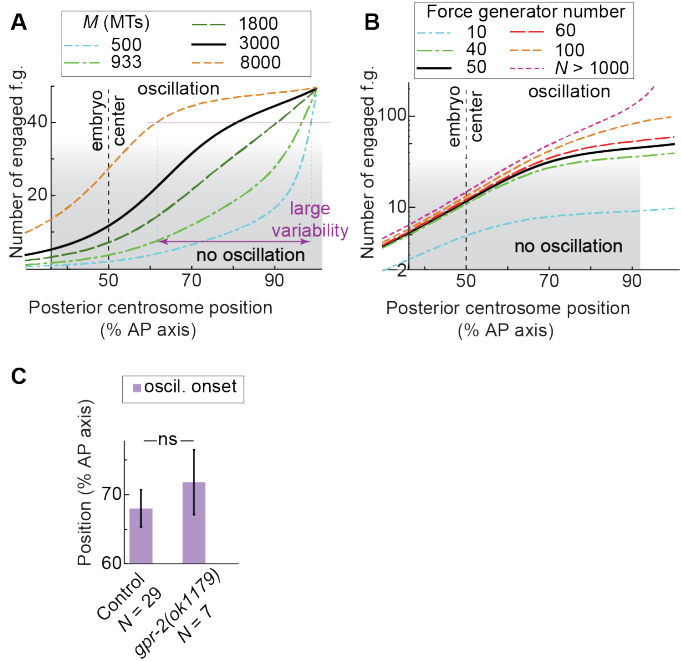

**Figure S2: Microtubule number affects the oscillation onset position, while force generator quantity above a threshold does not.**
(A-B) Modelled number of engaged force generators (f.g.) versus the posterior displacement of the centrosome along the anteroposterior (AP) axis as a percentage of the embryo length: variations (A) in total number of microtubules $M$ emanating from a single centrosome and (B) in total number of force generators $N$. In both cases, control values are black; green and blue are lower values; and red and orange are higher ones. The parameters used are listed in Table S5. Grey shading indicates when the number of engaged force generators was too low to permit oscillation (below threshold). Purple thin lines give a variability scale. (C) Posterior centrosome position at oscillation onset upon depletion of active force generators through mutation *gpr-2(ok1179)* ($N$ = 7 embryos) compared to $N$ = 29 control embryos at 18°C, both with GFP::γ-tubulin labelling of centrosomes. Error bars indicate SD and ns indicates no significant difference (Methods).

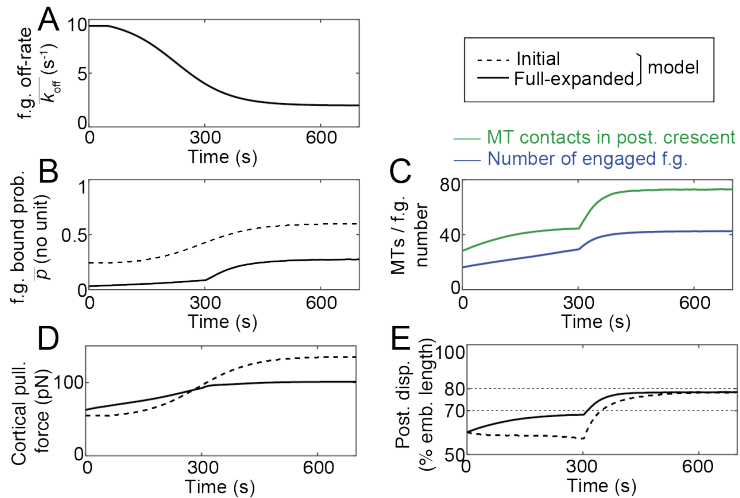

**Figure S3: Simulation of the displacement of the posterior centrosome using full-expanded model.**
Typical run showing: (A) the force generator (f.g.) detachment rate (inverse of processivity), which is the control parameter encoding the progression through mitosis (Pecreaux et al., 2006); (B) the probability $\bar{p}$ for a force generator to be pulling; (C) the number of astral microtubules (MTs) contacting the cortex in the posterior crescent (green) and the number of engaged f.g. in this same region (blue); (D) the cortical force in pN exerted by these force generators on the posterior centrosome projected on the anteroposterior (AP) axis; and (E) the posterior displacement subsequently obtained. The parameters used are listed in Table S5. Dashed lines represent the results of the initial model that only accounts for a temporal control of pulling forces, while solid lines correspond to the full-expanded one that combines the positional switch through microtubule dynamics on top of the temporal control through force generator dynamics (Suppl. Model).

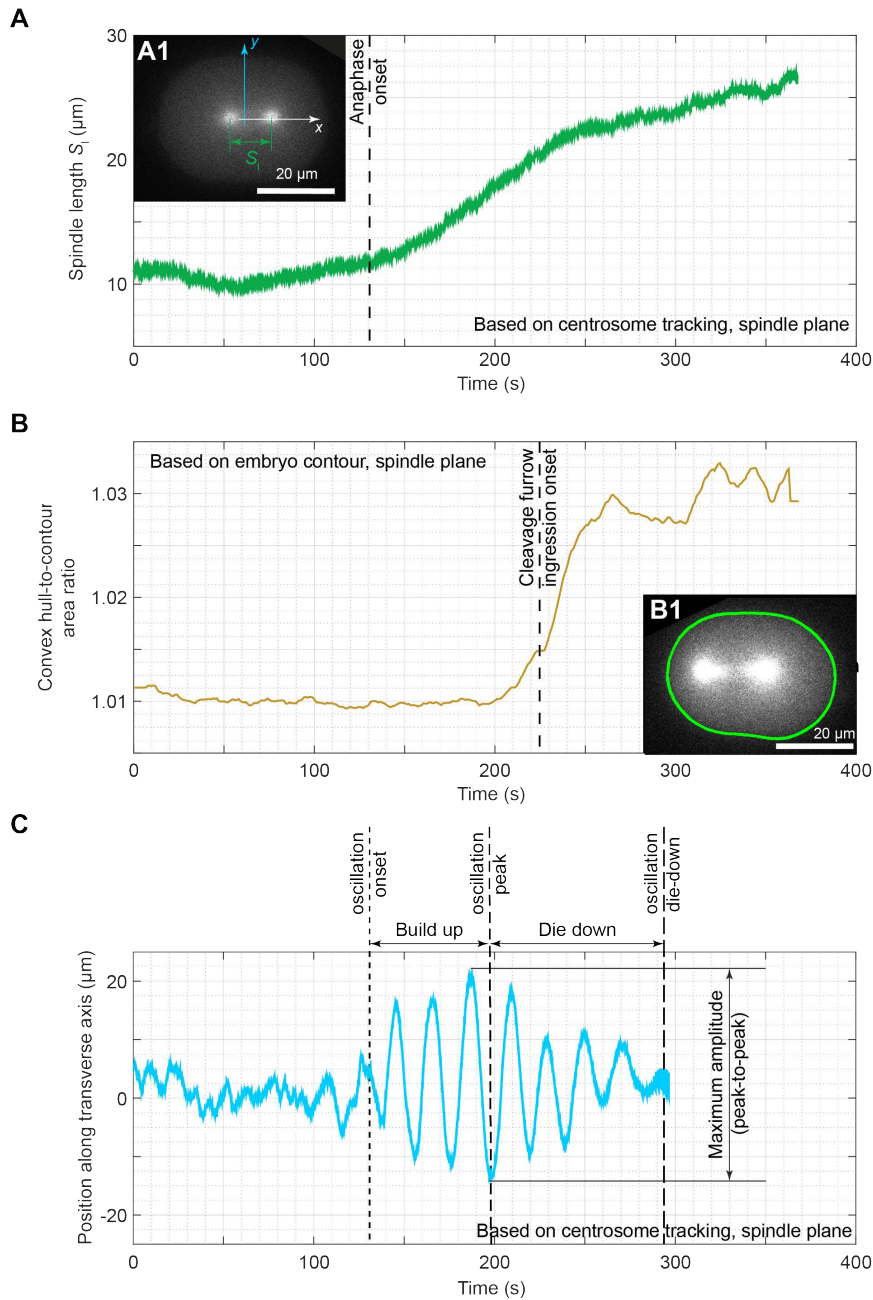

**Figure S4: Landmarks used to set reference times and oscillation characteristics.**
*C. elegans* centrosome trajectories and embryo shapes were measured in the spindle plane by viewing either the α-tubulin::YFP-labelled microtubule strain (inset B1) or the GFP::γ-tubulin-labelled centrosome strain (inset A1). (A) The spindle length was measured as the distance between the two centrosomes. Anaphase onset (dashed line) was defined as the inflection point towards a steeper increase in spindle length (Pecreaux et al., 2016). (B) The ratio of the convex contour to the real contour is used to measure the convexity of the embryo contour (inset B1, Suppl. Methods) and plotted across time. The cleavage furrow ingression timing (dashed line) is set as the inflection point in the convex hull-to-contour ratio steep increase. (C) Position of the posterior centrosome along the transverse axis of the

embryo. The timings for the oscillation onset, peak amplitude, and die-down are delineated. The vertical arrow indicates the maximum amplitude measurement (i.e. peak-to-peak distance in consecutive extrema).

| parameter | value | Description and estimate |
|---|---|---|
| | *Initial model parameters* | |
| $N$ | 50 (38 for the initial model) | Number of force generators per half cortex (Grill et al., 2003) (identical on both sides (Rodriguez Garcia et al., 2017)). $N = 38$ when simulating the initial tug-of-war to ensure a similar posterior displacement as in expanded model, all other parameters remaining similar. |
| $K$ | 5 µN/m | Centering spring stiffness. Same order of magnitude as (Garzon-Coral et al., 2016; Pecreaux et al., 2016). |
| $\Gamma$ | 140 µN.s/m | Damping due to microtubule network (Howard, 2006) and cytoplasm viscosity along transverse axis (Garzon-Coral et al., 2016). Used to model oscillations only. |
| $k_{on}^{post}$ | 3 s$^{-1}$ | Fixed force generator attachment rate (on-rate) on posterior side, used only in initial model. |
| $k_{on}^{ante}$ | 2 s$^{-1}$ | Fixed force generator attachment rate (on-rate) on anterior side, used only in initial model; its anterior value being lower than posterior to encode the polarity (Rodriguez Garcia et al., 2017). |
| $x_{ante}$ | 42% | Fixed anterior centrosome position (in % of embryo length), corresponding to experimental position at anaphase onset. |
| $f_c$ | 1.5 pN | Force generator detachment rate sensitivity to force (Pecreaux et al., 2006). |
| $f'$ | 3 µN.s/µm | Slope of the force generator force velocity relation (Pecreaux et al., 2006). |
| $\bar{f}$ | 6 pN | Force generator stall force (Pecreaux et al., 2006). |
| $\eta$ | $\langle |\eta| \rangle = 0$ $\langle |\eta|^2 \rangle = 0.1\delta(t)$ | Stochastic noise modelling the binding and unbinding of force generators (Pecreaux et al., 2006). |
| | *(Full) Expanded model parameters* | |
| $a$ | 24.6 µm | Embryo half-length (along AP axis), measured in this study. |
| $b$ | 15.75 µm | Embryo half-width (transverse to AP axis), measured in this study. |
| $M$ | 3000 | Number of microtubules emanating from each centrosome (O'Toole et al., 2003; Redemann et al., 2017). |
| $\alpha$ | 2.15 µm$^{-1}$ | Microtubule dynamics parameter, corresponding to known growing/shrinking rates (Kozlowski et al., 2007; Srayko et al., 2005) and cortex residency time $\tau = 1.25$ s (this work and (Kozlowski et al., 2007)). |
| $x_0^{post}$ | 70% | Position of the boundary of the active region in posterior embryo half (in % of embryo length) (Krueger et al., 2010). |

| | | Parameters for simulating posterior displacement |
|---|---|---|
| $\Gamma$ | 300 $\mu$N.s/m | Damping due to microtubule network (Howard, 2006) and cytoplasm viscosity along AP axis (Garzon-Coral et al., 2016). |
| $x_0^{ante}$ | 40% | Position of the boundary of active force generator region in anterior embryo half (in % of embryo length) based on force distribution (Grill et al., 2003). |
| $\lambda$ | 0.5 | Weakening factor of anterior forces to account for spindle pole uncoupling during elongation (Maton et al., 2015). |
| $\overline{k_{off}^0}$ | 4 s$^{-1}$ | Force generator off-rate at metaphase to anaphase transition, estimated from (Rodriguez Garcia et al., 2017). |
| $\overline{k_{off}^\infty}$ | 2 s$^{-1}$ | Final force generator off-rate (Rodriguez Garcia et al., 2017). |
| $\overline{k_{off}^{init}}$ | 10 s$^{-1}$ | Initial force generator off-rate. Time-dependent detachment rate (off-rate) $\overline{k_{off}(t)}$ varies following a sigmoid (Pecreaux et al., 2006). |
| $\beta_{post}$ | 15 | Affinity factor in posterior half of the cortex to account for the increased on-rate (Rodriguez Garcia et al., 2017). Set to have a number of active force generators in the range of 10 to 100 and twice larger in posterior than in anterior (Grill et al., 2003). |
| $\beta_{ante}$ | 7.5 | Affinity factor in anterior half of the cortex to account for the on-rate anteroposterior imbalance (Rodriguez Garcia et al., 2017). |
| $\widehat{k_{on}}$ | 0.025 s$^{-1}$ | Amplitude of the force generator attachment rate to the microtubule in the expanded model (Suppl. Model). |

**Table S5: Parameters used for modelling and simulations.**
Parameters used to calculate the number of engaged force generators with the expanded model and to simulate the posterior displacement using the full–expanded model. The estimates were similar to the initial model if applicable or based on experimental results cited in rightmost column.

**References**

Garzon-Coral, C., H.A. Fantana, and J. Howard. 2016. A force-generating machinery maintains the spindle at the cell center during mitosis. *Science*. 352:1124-1127.
Grill, S.W., J. Howard, E. Schaffer, E.H. Stelzer, and A.A. Hyman. 2003. The distribution of active force generators controls mitotic spindle position. *Science*. 301:518-521.
Howard, J. 2006. Elastic and damping forces generated by confined arrays of dynamic microtubules. *Phys Biol*. 3:54-66.
Kozlowski, C., M. Srayko, and F. Nedelec. 2007. Cortical microtubule contacts position the spindle in C. elegans embryos. *Cell*. 129:499-510.
Krueger, L.E., J.C. Wu, M.F. Tsou, and L.S. Rose. 2010. LET-99 inhibits lateral posterior pulling forces during asymmetric spindle elongation in C. elegans embryos. *J Cell Biol*. 189:481-495.
Maton, G., F. Edwards, B. Lacroix, M. Stefanutti, K. Laband, T. Lieury, T. Kim, J. Espeut, J.C. Canman, and J. Dumont. 2015. Kinetochore components are required for central spindle assembly. *Nat Cell Biol*. 17:697-705.
O'Toole, E.T., K.L. McDonald, J. Mantler, J.R. McIntosh, A.A. Hyman, and T. Muller-Reichert. 2003. Morphologically distinct microtubule ends in the mitotic centrosome of Caenorhabditis elegans. *J Cell Biol*. 163:451-456.

Pecreaux, J., S. Redemann, Z. Alayan, B. Mercat, S. Pastezeur, C. Garzon-Coral, A.A. Hyman, and J. Howard. 2016. The Mitotic Spindle in the One-Cell C. elegans Embryo Is Positioned with High Precision and Stability. *Biophys J*. 111:1773-1784.

Pecreaux, J., J.C. Roper, K. Kruse, F. Julicher, A.A. Hyman, S.W. Grill, and J. Howard. 2006. Spindle oscillations during asymmetric cell division require a threshold number of active cortical force generators. *Curr Biol*. 16:2111-2122.

Redemann, S., J. Baumgart, N. Lindow, M. Shelley, E. Nazockdast, A. Kratz, S. Prohaska, J. Brugues, S. Furthauer, and T. Muller-Reichert. 2017. C. elegans chromosomes connect to centrosomes by anchoring into the spindle network. *Nature communications*. 8:15288.

Rodriguez Garcia, R., L. Chesneau, S. Pastezeur, J. Roul, M. Tramier, and J. Pecreaux. 2017. Dynein dynamics at the microtubule plus-ends and cortex during division in the C. elegans zygote. *bioRxiv*:118604.

Srayko, M., A. Kaya, J. Stamford, and A.A. Hyman. 2005. Identification and characterization of factors required for microtubule growth and nucleation in the early C. elegans embryo. *Dev Cell*. 9:223-236.

# Supplementary method to "LET-99-dependent spatial restriction of active force generators makes spindle's position robust"

## "Landing" assay: Pipeline to measure microtubule contact densities and dynamics at the cortex

*Choosing the labelling*

Two main strategies could be envisaged to investigate microtubule dynamics (Gierke et al., 2010): the labelling of microtubule growing ends with EB homolog proteins and the labelling of the entire microtubule using tubulin tagged with fluorescent proteins. Here, we chose the second option by using nematode strains with YFP::α-tubulin and GFP::β-tubulin transgenes for *C. elegans* and *C. briggsae* embryos, respectively. It allowed us to measure the duration of the residency of microtubules at the cortex, disregarding whether they were growing, shrinking or clamped in any of these two states. Dynein, which causes the pulling forces that position the spindle (Nguyen-Ngoc et al., 2007), was indeed reported to also cause such a "residing state" *in vitro* (Laan et al., 2012). We could then relate microtubule cortical contacts and spindle positioning, in particular posterior displacement during anaphase. In this perspective, we checked that the $\alpha$-tubulin overexpression in the *C. elegans* strain used for the "landing" assay was not resulting in any significant phenotypes by investigating the trajectories of the centrosomes and their oscillations, a sensitive read-out of the pulling forces (Pecreaux et al., 2006a). To do so, we compared YFP::α-tubulin microtubule-labelled strain with GFP::γ-tubulin centrosome-labelled strain. We observed significant difference neither in the posterior and anterior centrosome trajectories along the anteroposterior (AP) axis throughout mitosis (Figure S1A-B), nor in posterior centrosome oscillation maximum amplitude or timings (Table below). We concluded that the α-tubulin labelling does not perturb the centrosomal oscillation phenotype.

| Characteristic measured (mean ± s.e.m.) | γ-tubulin labelling (*N* = 39, 23ºC) | α-tubulin labelling (*N* = 8, 23ºC) *p* for Student's t-test |
|---|---|---|
| Maximum oscillation amplitude (posterior) (% of embryo width) | 19.43 ± 0.09 | 16.79 ± 0.51 *p* = 0.088 |
| Maximum oscillation amplitude (posterior) (μm) | 6.14 ± 0.03 | 5.54 ± 0.17 *p* = 0.224 |
| Oscillation die-down (s, from anaphase onset) | 116.5 ± 0.3 | 115.08 ± 1.72 *p* = 0.774 |
| Oscillation duration (s, from anaphase onset) | 121.62 ± 0.49 | 121.88 ± 3.79 *p* = 0.980 |
| Maximum oscillation peak (posterior) (s, from anaphase onset) | 47.1 ± 0.3 | 43.36 ± 3.50 *p* = 0.688 |
| Embryo length (μm) | 49.23 ±0.07 | 48.23 ± 0.41 *p* = 0.39 |
| Embryo width (μm) | 31.53 ± 0.03 | 33 ± 0.14 *p* = 0.003 |

*Imaging of microtubule contacts at the cortex*

We imaged *C. elegans* or *C. briggsae* one-cell embryos at the cortex plane in contact with the glass slide (Figure 1D), viewing from the nuclear envelope breakdown (NEBD) until the end of cell division. We did our utmost to preserve the embryo shapes. The thickness of the perivitelline space (Olson et al., 2012) therefore meant we had to use spinning disk microscopy rather than TIRF. Cortical microtubule contact tracking was thus performed on a LEICA DMI6000 / Yokogawa CSU-X1 M1 spinning disc microscope, using a HCX Plan Apo 100x/1.4 NA oil objective. Illumination was performed using a white-light Fianium laser filtered around 514 nm in a homemade setup (Roul et al., 2015), sold by Leukos (Limoges, France). To account for the fast speed of microtubule dynamics at the cortex, images were acquired at an exposure time of 100 ms (10 Hz) using an ultra-sensitive Roper Evolve EMCCD camera and the MetaMorph software (Molecular Devices) without binning. During the experiments, the embryos were kept at 23°C. To image embryos at the cortex, we typically moved the focus to 12 to 15 µm below the spindle plane (Figure 1D).

*Preprocessing of the cortical images*

Since microtubule tubulin spot signals were very weak at the cortex (Figure 1A), we denoised the images to increase the signal-to-noise ratio (Figure 1B). This noise reduction usually relies on the assumption that the noise is non-correlated in space and time and that it follows a Gaussian or Poisson distribution over space or time. We opted for Kalman filtering/denoising (Kalman, 1960), setting the gain to 0.5, and the initial noise estimate to 0.05 performing in time and not requiring hypothesis in space.

*Automated tracking of YFP::α-tubulin fluorescent spots at the cortex*

Because of the low signal-to-noise ratio and the large quantity of tracks present at the cortex, we looked for an algorithm with powerful track detection and segment linking capabilities. We opted for u-track (Jaqaman et al., 2008), using the parameters below:

| **Detection** | |
|---|---|
| Gaussian standard deviation | 1.4 |
| Alpha-value for initial detection of local maxima | 0.08 |
| Rolling window time-averaging | 3 |
| Iterative Gaussian mixture-model fitting | 0 |
| **Tracking** | |
| Maximum gap to close | 5 |
| Merge split | 0 |
| Minimum length of track segments from first step | 3 |
| **Cost function frame-to-frame linking** | |

| | |
|---|---|
| Flag for linear motion | 1 |
| Allow instantaneous direction reversal | 1 |
| Search radius lower limit | 1 |
| Search radius upper limit | 3 |
| Standard deviation multiplication factor | 3 |
| Nearest neighbour distance calculation | 1 |
| Number of frames for nearest neighbour distance calculation | 4 |
| **Cost function close gaps** | |
| Flag for linear motion | 1 |
| Search radius lower limit | 1 |
| Search radius upper limit | 3 |
| Standard deviation multiplication factor | 3 |
| **Nearest neighbour distance calculation** | |
| Number of frames for nearest neighbour | 4 |
| Penalty for increasing gap length | 1.5 |
| Maximum angle between linear track segments | 30 |

*Measuring microtubule residency time at the cortex and consistency check*
We calculated the histogram for the durations of microtubule contacts at the cortex. We used a bin size of 100 ms, equal to the image acquisition time. The exponential fit of the histogram yielded a microtubule lifetime (Figure 1E), consistent with previous work (Kozlowski et al., 2007). Indeed, the duration of a microtubule contact at the cortex is limited by its switching to depolymerisation (catastrophe), which is a first order stochastic process. Furthermore, the increasing count of microtubule cortical contacts at anaphase (Figure 1FG, Figure 2A,D) was consistent with the increased microtubule nucleation rate previously reported (Srayko et al., 2005).

*Validation of the microtubule contact analysis at the cortex*
To gain confidence on the processing beyond the consistency, we validated the u-track parameters by analysing fabricated fluorescence images of known dynamics (Figure S1C-E), which mimic our cortical images (Costantino et al., 2005). In further details, we simulated stochastic trajectories of particles (Figure S1C) that displayed a limited random motion: $x_{i,j}(t+1) = x_{i,j}(t) + \xi\sqrt{2Dt}$, where $x_{i,j}(t)$ represents the coordinates in two dimensions at time $t$, $\xi$ is a random number and $D$ is the diffusion

3

coefficient. The duration of the tracks (length) was sampled from an exponential distribution. The intensity was set similar to experimental ones and encoded by the quantum yield parameter (Qyield). We then plotted the instantaneous positions and applied a Gaussian blur filter to mimic the effect of the point-spread function in fluorescence microscopy. We mimicked the background noise by adding at each pixel a sampling of a Gaussian distribution normalized to ε, with formula reading $A_{noisy} = A + \varepsilon M$ and corresponding to a signal-to-noise ratio $max(A)/\varepsilon$ (Figure S1D). This simulation provided a realistic scenario to test the image-processing pipeline. Details of the parameters used for simulation can be found below. We were able to get very good matching of duration distributions either plotting parameters used to generate the simulated tracks and the ones recovered through analysing fabricated images (Figure S1E). This validated our microtubule contact tracking approach.

| Image size | 256 x 256 |
|---|---|
| Duration | 3000 frames (300 s) |
| Density of particles (tracks) | 1 particle /µm$^2$ |
| QYield | 0.61 |
| Pixel size | 0.139 nm |
| Sampling rate | 0.1 s |
| PSF type | Gaussian |
| PSF size | 0.244 µm |
| Bits | 12 |
| Diffusion coefficient | 0.002 µm$^2$/s |
| Background noise standard deviation | 0.5 |
| Mean track lifetime | 1 s |

*Computing microtubule contact densities at the cortex ("landing" assay)*
The "landing" assay consisted of measuring the cortical microtubule contacts (Figure 1D) during the different phases of mitosis. The part of the embryo contacting the cover-slip was divided into ten regions of equal width along its long axis (AP axis). The u-track algorithm allowed us to follow the microtubule contacts at the cortex frame-by-frame, and to have access to their trajectories (Jaqaman et al., 2008). We segmented the embryo cytoplasm so we could track embryo shape changes throughout cell division using the active contour algorithm (Pecreaux et al., 2006b), thus obtaining data on how the lengths and areas of the embryos evolved during mitosis. We were then able to count the microtubules contacting the cortex in the ten regions along the embryo length. To reinforce our findings, the distribution of the microtubule contacts was averaged over time for 10 s. Finally, we used the onset of cytokinesis cleavage furrow ingression as a temporal reference for the alignment of the different microtubule cortical contact density maps, and then averaged the embryos to get the final density map.

*Timing of cytokinesis cleavage furrow ingression onset to overlay centrosome trajectories to microtubule cortical contact density*

We determined the timing of the onset of cleavage furrow ingression in both planes by taking advantage of the dyed cytoplasmic fraction to detect the embryo contours. These contours were obtained via the active contour algorithm (Pecreaux et al., 2006b). We then set the onset of cleavage furrow ingression to equal the fast increase in embryo shape ratio of the convex area to the active contour one (Figure S4B), which in practice meant when it grew above 1.012. In the mid-plane, we calibrated the average time between anaphase (starting at the inflection point during spindle elongation (Pecreaux et al., 2016) (Figure S4A)), and the onset of cleavage furrow ingression. We then used this to estimate anaphase onset from cortical cleavage furrow ingression onset measurements, matching the timing from the "landing" and "centrosome-tracking" assays to plot overlays (Figures 2A,D).

**References**

Costantino, S., J.W. Comeau, D.L. Kolin, and P.W. Wiseman. 2005. Accuracy and dynamic range of spatial image correlation and cross-correlation spectroscopy. *Biophys J*. 89:1251-1260.

Gierke, S., P. Kumar, and T. Wittmann. 2010. Analysis of microtubule polymerization dynamics in live cells. *Methods in cell biology*. 97:15-33.

Jaqaman, K., D. Loerke, M. Mettlen, H. Kuwata, S. Grinstein, S.L. Schmid, and G. Danuser. 2008. Robust single-particle tracking in live-cell time-lapse sequences. *Nat Methods*. 5:695-702.

Kalman, R.E. 1960. A New Approach to Linear Filtering and Prediction Problems. *Journal of Basic Engineering*. 82:35-45.

Kozlowski, C., M. Srayko, and F. Nedelec. 2007. Cortical microtubule contacts position the spindle in C. elegans embryos. *Cell*. 129:499-510.

Laan, L., N. Pavin, J. Husson, G. Romet-Lemonne, M. van Duijn, M.P. Lopez, R.D. Vale, F. Julicher, S.L. Reck-Peterson, and M. Dogterom. 2012. Cortical dynein controls microtubule dynamics to generate pulling forces that position microtubule asters. *Cell*. 148:502-514.

Nguyen-Ngoc, T., K. Afshar, and P. Gonczy. 2007. Coupling of cortical dynein and G alpha proteins mediates spindle positioning in Caenorhabditis elegans. *Nat Cell Biol*. 9:1294-1302.

Olson, S.K., G. Greenan, A. Desai, T. Muller-Reichert, and K. Oegema. 2012. Hierarchical assembly of the eggshell and permeability barrier in C. elegans. *J Cell Biol*. 198:731-748.

Pecreaux, J., S. Redemann, Z. Alayan, B. Mercat, S. Pastezeur, C. Garzon-Coral, A.A. Hyman, and J. Howard. 2016. The Mitotic Spindle in the One-Cell C. elegans Embryo Is Positioned with High Precision and Stability. *Biophys J*. 111:1773-1784.

Pecreaux, J., J.C. Roper, K. Kruse, F. Julicher, A.A. Hyman, S.W. Grill, and J. Howard. 2006a. Spindle oscillations during asymmetric cell division require a threshold number of active cortical force generators. *Curr Biol*. 16:2111-2122.

Pecreaux, J., C. Zimmer, and J.C. Olivo-Marin. 2006b. Biophysical active contours for cell tracking I: Tension and bending. *In* 2006 Ieee International

Conference on Image Processing, Icip 2006, Proceedings, October 8-11 2006, Atlanta, GA, USA. 1949-1952.

Roul, J., J. Pecreaux, and M. Tramier. 2015. Method for controlling a plurality of functional modules including a multi-wavelength imaging device, and corresponding control system. Vol. WO2015144650 A1. U.D.R. Cnrs - Centre National De La Recherche Scientifique, editor.

Srayko, M., A. Kaya, J. Stamford, and A.A. Hyman. 2005. Identification and characterization of factors required for microtubule growth and nucleation in the early C. elegans embryo. *Dev Cell*. 9:223-236.


\title{\vspace{-2cm}{Supplementary model to ``LET-99-dependent spatial restriction of active force generators makes spindle's position robust."}}


\author[1,2]{H. Bouvrais\thanks{
           To whom correspondence should be addressed: helene.bouvrais@univ-rennes1.fr, jacques.pecreaux@univ-rennes1.fr}\textsuperscript{,}}
\author[1,2]{L. Chesneau}
\author[1,2]{S. Pastezeur}
\author[3]{M. Delattre}
\author[1,2]{J. P\'ecr\'eaux\samethanks\textsuperscript{,}}
\affil[1]{CNRS UMR 6290,
	    F-35043  Rennes,\space France.}
\affil[2]{University of Rennes 1, UEB, SFR Biosit, School of Medicine, F-35043 Rennes, France}
\affil[3]{Laboratory of Molecular Biology of the Cell, \'{E}cole Normale Sup\'{e}rieure de Lyon, CNRS, F-69363 Lyon, France}

\maketitle

\renewcommand\thefootnote{\arabic{footnote}}


\tableofcontents
\newpage


\section{Introduction}

During the division of nematode zygote, the spindle undergoes a complex choreography. Firstly, during prophase, the pronuclei-centrosomes complex (PCC) moves from posterior half of the embryo to its center, during the so-called centring phase \citep{ahringer03} and concurrently the two centrosomes align along the antero-posterior-axis. We previously found, in contrast to \textit{Caenorhabditis elegans} embryo, that this displacement was a bit excessive in \textit{C. briggsae} reaching a slightly more anterior position, a phenomenon called \textit{overcentration} \citep{kimura07,riche13}. The consequence is a delay in spindle posterior displacement for this species with respect to \textit{C. elegans}. Interestingly, the same proteins that cause the anaphase posterior displacement are needed for this \citep{riche13}, namely the trimeric compex GPR-1/2$^\text{LGN}$, LIN-5$^\text{NuMA}$ and dynein \citep{nguyen07}. Later on, during prometaphase and metaphase, the spindle is maintained in the middle by centring forces that are independent of GPR-1/2$^\text{LGN}$ and may be caused by microtubule pushing on the cell cortex \citep{pecreaux16}. Finally, during late metaphase and anaphase, GPR-1/2-dependent cortical pulling forces become dominant and displace the spindle posteriorly, make it oscillate, and contribute to its elongation \citep{grill03,labbe04,pecreaux06}. 

We aim here to complement our previously published ``tug-of-war" model \citep{grill05,pecreaux06}, later called initial model, which was mainly focused on the dynamics of cortical force generators (f.g.), by including the dynamics of astral microtubules (MTs). Indeed, we mapped the microtubule contacts at the cortex and revealed that they mostly concentrated in cortical regions close to the centrosomes \citep{bouvrais18}. In consequence, the position of the centrosomes, as microtubule organizing centres (MTOC), regulates the quantity of engaged force generators pulling on astral microtubules and in turn spindle's anaphase oscillation and posterior displacement. \\

First, focusing on the oscillation onset, we expanded our initial model of spindle oscillation to account for microtubule dynamics. We detailed the expanded model and then explored how this novel positional regulation combines with the one by force generator processivity previously reported \citep{pecreaux06}. Second, through a stochastic simulation approach, we looked at the feedback loop created between the position of the posterior centrosome and the pulling forces contributing to spindle displacement. \\

\section{Modelling the positional switch on oscillation onset}

\subsection{Quantity of microtubules reaching the posterior crescent of active force generators}

Recent work suggested that force generators would be active only on a posterior cap instead of the whole posterior half cortex of the embryo \citep{krueger10}. This means that only the microtubules hitting the posterior crescent of the cortex would contribute to spindle displacement by binding to active force generators. We thus calculated the number of microtubules reaching this so-called active region of the cortex. 

\subsubsection{Modelling hypotheses and microtubule dynamics parameter estimates} \label{MT_estimates}

We set to explore whether the number of microtubules reaching the cortex, assumed to be in excess during anaphase \citep{grill05,pecreaux06}, could be limiting prior to oscillation onset. Key to assess this possibility was an estimate of the total number of microtubules and their dynamics. Based on previously published experiments, we assessed the following microtubule related parameters:
\begin{itemize}
\item \textbf{Total number of microtubules.}
To assess the number of microtubule nucleation sites at the centrosome (CS), we relied on electron microscopy images of the centrosomes \citep{redemann16}, which suggested $3000$ or more microtubules emanating per centrosome. This order of magnitude was previously proposed by O'Toole and collaborators \citep{otoole03}. More specifically in the figure 3, authors provide a slice of about $0.85\;\mu\text{m}$ thick (as estimated from video 8) displaying 520 astral microtubules, while centrosome diameter was estimated to  $1.5\;\mu\text{m}$. Only a slice of centrosome was viewed in this assay, so that the number of microtubule nucleation sites per CS was extrapolated to a least $1800$ considering the centrosome as a whole sphere. In this work, we set the number of microtubules to $M=3000$. Variation of this number within the same order of magnitude does not change our conclusions.

\item \textbf{The microtubules are distributed around each centrosome in an isotropic fashion.}
We hypothesized an isotropic distribution of microtubules around each centrosome following \citep{howard06}. This was also suggested through electron microscopy \citep{redemann16}.

\item \textbf{Free-end catastrophes are negligible.}
With the above estimate of the microtubule number and considering a microtubule growing speed in the cytoplasm $v^+ = 0.67\;\mu\text{m/s}$ \citep{srayko05} and a shrinking one $v^- = 0.84\;\mu\text{m/s}$ \citep{kozlowski07}, we could estimate that about 70 microtubules reach the cell periphery (assumed to be at $15\;\mu\text{m}$ from the centrosome) at any moment and per centrosome, if the free-end catastrophe rate is negligible. This estimate appears consistent with the instantaneous number of force generators in an half-cortex, estimated between 10 and 100 \citep{grill03}. 

Furthermore, it was recently proposed that the catastrophe rate could be as high as $0.25\;\text{s}^{-1}$ in the mitotic spindle \citep{redemann16}. On the one hand, this might be specific to this organelle since the spindle is much more crowded than the cytoplasm. On the other hand, these authors proposed a total number of microtubules two to three folds larger than our estimate. We asserted that our conservative estimate of the microtubule quantity combined with the negligible free-end catastrophe resulted in similar modelling results, with the advantage of the simplicity over a full astral microtubule model. In other words, we focused on the fraction of astral microtubules not undergoing free-end catastrophe, which was the only one measurable at the cortex. 

We next wondered whether the assumption of negligible free-end catastrophe is consistent with our measurement of microtubule contact density at the cortex. After \citep{redemann16}, the vast majority of microtubules emanating from the centrosome are astral: we thus assumed that the kinetochore and spindle microtubules were negligible in this estimate. Focusing on metaphase and with a residency time of microtubule ends at the cortex $\tau = 1.25\;\text{s}$ \citep{kozlowski07,bouvrais18}, this led to about 100 microtubules contacting the cortex per centrosome, at any given time. Using our "landing" assay \citep{bouvrais18}, we could estimate the number of contacts in the monitored region at any given time to 5 microtubules. Extrapolating this to a whole centrosome and assuming the isotropic distribution of astral microtubules (\S \ref{sensitivity}), we found 26 cortical contacts of microtubules at any time in metaphase. Although a bit low, likely because of the conservative parameters of the image processing that could led to missing some microtubules, this experimental assessment was consistent with the theoretical estimate based on our hypotheses. Furthermore, it was also consistent with the measurement done by  \citep{garzon16}. In contrast, a non negligible catastrophe rate would have dramatically reduced that number of contacts at any given time. We concluded that free-end catastrophe rate was safely negligible.

\item \textbf{No microtubule nucleation sites are left empty at the centrosomes}
This is a classic hypothesis \citep{howard06}, recently supported by electron microscopy experiments \citep{redemann16}.

\end{itemize}

\subsubsection{Microtubule dynamics ``measure" the centrosome--cortex distance. \label{sensitivity}}

\paragraph{Probability for a microtubule to be at the cell cortex} Because microtubules spend most of their ``lifespan" growing to and shrinking from the cortex, the distance between the centrosome and the cortex limits the number of microtubules residing at the cortex at any given time. We could thus summarize microtubule dynamics in a single parameter $\alpha$ by writing the fraction of time spent by a microtubule at the cell cortex:

\begin{equation}
q\;=\;\frac{\tau}{\frac{d}{  v^+} + \frac{d}{ v^-} + \tau}\; =\; \frac{1}{1+\alpha d} \quad \text{with }\alpha=\frac{1}{v^+\tau} + \frac{1}{v^-\tau},
\end{equation}
where $d$ is the distance from the centrosome (MTOC) to the cortex (estimated to typically $d=15\;\mu\text{m}$, \textit{id est} about half of the embryo width). We then found $\alpha = 2.15\times10^6\;\text{m}^{-1}$ using the above microtubule dynamics parameters. 
This meant that the microtubule spent $q=3\%$ of its time at the cortex and the remaining time growing and shrinking. This fraction of time spent residing at the cortex was consistent with the estimate coming from investigating the spindle centering maintenance during metaphase \citep{pecreaux16}. 

\paragraph{Range of variations in the microtubule contact densities at the cortex.} 
The nematode embryo shape is close to an ellipsoid. Therefore, the centrosome displacement can vary the centrosome-cortex distance by 1.5 to 2 fold. We wondered whether the microtubule dynamics were so that one could observe significant variations in cortical microtubule-residing probabilities $q$. We estimated this sensitivity through the ratio $\rho$ of the probability of reaching the cortex when the centrosome was at its closest position $d_1$ (set to half of the embryo width, i.e. the ellipse short radius) divided by the probability when it was at its furthest position $d_2$ (chosen as half of the embryo length, i.e. the ellipse long axis).
\begin{equation}
\rho =\frac{1+\alpha d_2}{1+\alpha d_1}
\end{equation}
This curves had a sigmoid--like shape with $\lim\displaylimits_{\alpha \to 0}\rho=1$ and $\lim\displaylimits_{\alpha \to \infty}\rho= d_2 \left/ d_1 \right.$. 

\begin{figure}
   \begin{center}
      \includegraphics*[width=85mm]{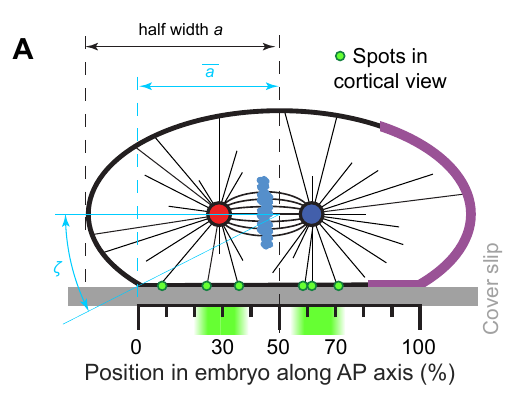}
   \end{center}
      \caption{\textbf{Experimental setup for viewing microtubule (MT) contact density at the cell cortex.} The scale represents the 10 regions along the anteroposterior (AP) axis used for analysis \citep{bouvrais18}. Red and blue disks represent the anterior and posterior centrosomes, respectively, and the light blue clouds are the chromosomes. Microtubules emanating from the centrosomes are thin black lines. The posterior crescent where the active force generators are located is the purple cortical region.}
      \label{fig1}
\end{figure}

Using our measurement of microtubule contact distribution at the cortex \citep{bouvrais18}, we calculated an experimental estimate of this sensitivity parameter, $\rho_{exp}\simeq 2$. 
On model side, because the experimental ``landing" assay did not enable us to view the very tip of the embryo (Figure \ref{fig1}), we compared the sensitivity ratio calculated from the density map with a theoretical one that did not use the half embryo length as maximum distance but the largest distance effectively measurable. For untreated embryos viewed at the spindle plane, the measured embryo length was $2a=49.2\;\mu\text{m}$, while imaging at the cortex, the length along anteroposterior (AP) axis (denoted with bars) was $2\bar{a}=38.0\;\mu\text{m}$ for the adhering part to the coverslip. We could calculate the truncation of the ellipse due to the adhesion through the polar angle $\zeta=\arccos \left( \bar{a}\left/ a \right. \right)$ of the boundary of the adhering region. We obtained $\zeta=39.4\degree$ which corresponded to a spindle plane to flattened cortex distance of $10\;\mu\text{m}$, using a parametric representation of the ellipse. During metaphase (set as the two minutes preceding anaphase onset), when the spindle is roughly centered \citep{pecreaux16}, the average spindle length was $11.8\;\mu\text{m}$ ($N$ = 8 embryos). The furthest visible region was thus at $d_2= 16.5\;\mu\text{m}$ while the closest one was at  $d_1=10\;\mu\text{m}$, leading to a sensitivity ratio $\rho=1.62$ consistent with the microtubule cortical contact density ratio observed {\it in vivo} for {\it C. elegans}. We concluded that microtubule dynamics in {\it C. elegans} enable the read-out of the posterior centrosome position through the probability of microtubules to be in contact with the cell cortex.   

\subsubsection{Number of microtubules reaching the cortex} 

\begin{figure}
   \begin{center}
      \includegraphics*[width=160mm]{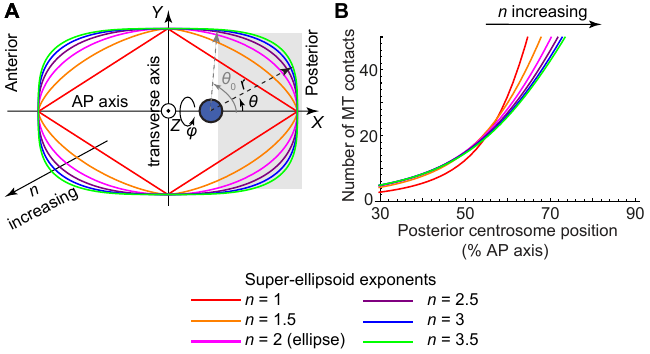}
   \end{center}
      \caption{\textbf{The regime change in the microtubule contacts is preserved after super-ellipsoid modelling of embryo shape using various exponents.} An embryo shape was modelled by super-ellipsoids of the variable $n$ (long axis $24.6\;\mu\text{m}$ and short axis $15.75\;\mu\text{m}$), with $n = 2$ representing the ellipse. (A) Super-ellipses with the exponent $n$ set to 1 to 3.5. The centrosome was positioned at 67\% of the anteroposterior (AP) axis and is a blue disk. Cartesian axes $(X,Y,Z)$ are indicated as well as spherical coordinates centred on the centrosome. The active region is grey, and its boundary was set to 70\% of embryo length and at angle in centrosome-centered spherical coordinates. (B) Number of microtubules contacting the active region versus the position of the posterior centrosome along the AP axis, which is shown as a percentage of embryo length. Embryo shape was modelled using super-ellipsoids of revolution based on the super-ellipses plotted in A, and the parameters used are listed in \citep[Table S5]{bouvrais18}.}
      \label{fig2}
\end{figure}

We set to estimate the variation of the total number of astral microtubule contacts emanating from a single centrosome versus the position of this centrosome along the AP axis. We worked in spherical coordinates $(r,\theta,\phi)$ centered on the posterior centrosome that displayed a slow posterior displacement assumed to be a quasi-static motion, with zenith pointing towards posterior. We denoted $\theta$ the zenith angle and $\phi$ the azimuth (Figure \ref{fig2}A).  We calculated the probability of a microtubule to reach the cortex in the active region, represented as $\theta\in [0,\theta_0 ]$ and $\phi\in [0,2\pi [$. We integrated over the corresponding solid angle and the number of microtubules reaching the cortex $\mathcal{M}\left(\mathcal{S},\alpha\right)$ came readily (Figure \ref{fig2}B):

\begin{eqnarray}
p\left(\mathcal{S},\alpha, \theta , \phi \right) & = & \frac{1}{1+\alpha\, r_\mathcal{S}\left(\theta,\phi\right)} {\color{blue} \operatorname{sin}\left(\theta\right)} \label{eq_p} \\
P\left(\mathcal{S},\alpha\right) & = & \int_{\theta=0}^{\theta_0} \int_{\phi=0}^{2\pi}  p\left(\mathcal{S},\alpha, \theta , \phi \right)    \operatorname{d}{\phi}\operatorname{d}\theta \label{eq_P} \\[15pt]
\mathcal{M}\left(\mathcal{S},\alpha\right) & = & M \left/ 4\pi \right. P\left(\mathcal{S},\alpha\right) \label{Mcal}
\end{eqnarray}
where $r_\mathcal{S}\left(\theta,\phi\right)$ is the distance centrosome--cortex in polar coordinates centered on the centrosome, dependent upon the shape of the cortex $\mathcal{S}$ and the boundary of the active force generator region  $\theta_0$ (Figure \ref{fig1}). We observed a switch-like behaviour as the posterior centrosome went out of the cell centre and closer to the posterior side of the embryo (Figure \ref{fig2}B).

\subsection{Towards the expanded tug-of-war model \label{EToW}}

In the initial model \citep{grill05,pecreaux06}, we made the assumption that the limiting factor was the number of engaged cortical force generators while in comparison, the astral microtubules were assumed to be in excess. It resulted that oscillations were driven by the force generator quantity and dynamics.  In the linearised version of the initial model, the persistence of force generators to pull on microtubules (i.e. their processivity) mainly governed the timing and frequency of the oscillations, while the number of force generators drove the amplitude of oscillations \citep{pecreaux06}. However, since the number of microtubules reaching the cortex could be limiting \citep{kozlowski07}, we expanded the initial model of anaphase oscillations to account for this possible limitation.

\subsubsection{The initial model \label{orig}}

We provide here a brief reminder of the initial tug-of-war model \citep{pecreaux06}.  
It featured cortical force generators exhibiting stochastic binding to and detaching from microtubules at rates $k_{on}$ and $k_{off}$ ($\overline{k_{off}}$ being the detachment rate at stall force $\bar{f}$), respectively. The force generators were assumed to act close to stall force. The mean probability for a force generator to be pulling on a microtubule then reads $\bar{p}=k_{on}\left/ \left( k_{on} + \overline{k_{off}} \right)\right.$. The active force generators were distributed symmetrically between the upper and lower posterior cortices but asymmetrically between anterior and posterior cortices \citep{grill03}. In the model, we also included two standard properties of the force generators: firstly, a force-velocity relation $f=\bar{f}-f^\prime\,v$, with $f$ the current force, $v$ the current velocity, and $f^\prime$ the slope of the force velocity relation ; secondly, a linearised load dependent detachment rate $k_{off} = \overline{k_{off}}\left(1 -\frac{\bar{f}-f}{f_c}\right)$, with $f_c$ the sensitivity to load/pulling force, assuming that force generator velocity was low, i.e. they acted close to the stall force \citep{pecreaux06}. We finally denoted $\Gamma$ the passive viscous drag, related in part to the spindle centring mechanism \citep{garzon16,pecreaux16,howard06} and $\bar{N}$ the number of available force generators in the posterior cortex. 

A quasi-static linearised model of the spindle posterior displacement reads:
\begin{equation}
I(\bar{p})\ddot{y}+\left(\Gamma - \Xi(\bar{p}) \right) \dot{y} + Ky = 0,
\label{eqToW}
\end{equation}
with
\begin{equation}
\Xi(\bar{p}) = 2\bar{N}\left\{ \frac{\bar{f}}{f_c}\bar{p}\left[\left( 1-\bar{p}\right) - \frac{f_c}{\bar{f}}\right]\right\}f^\prime,
\label{eqXi}
\end{equation}
and
\begin{equation}
I(\bar{p}) = 2\bar{N}\left\{ \frac{\bar{f}}{f_c}\bar{p}\left( 1-\bar{p}\right)\right\}f^\prime\frac{\bar{p}}{k_{on}},
\label{eqI}
\end{equation}
with $K$ the centering spring stiffness and $I$ the inertia resulting from stochastic force generator binding and unbinding.
The spindle oscillations develop when the system becomes unstable, meaning when the negative damping $\Xi(\bar{p})$ overcomes the viscous drag $\Gamma$.

\subsubsection{Evolution of the initial model to account for the polarity encoded through force generator on-rate \label{orig_kon}}

When we designed the initial model, it was known that the spindle posterior displacement was caused by an imbalance in the number of active force generators \citep{grill03}, i.e. the number of force generators engaged in pulling on astral microtubules or ready to do so when meeting an astral microtubule. However, the detailed mechanism building this asymmetry was elusive. We recently investigated the dynamics of dynein at the cell cortex \citep{rodriguez17} and concluded that the force imbalance rather resulted from an asymmetry in force generator attachment rate to the microtubule. This asymmetry reflects the asymmetric location of GPR-1/2 \citep{park08,riche13}. More abundant GPR-1/2 proteins at posterior cortex could displace the attachment reaction towards more binding/engaging of force generators. Therefore, to simulate the posterior displacement of the posterior centrosome (\S \ref{simulPostDisp}), we rather used the equations above (Eq. \ref{eqToW}-\ref{eqI}) with distinct on-rates between anterior and posterior sides and equal quantity of available force generators.

\subsubsection{Number of engaged force generators: modelling the binding of a microtubule to a force generator \label{sec_Ka}}
\paragraph{Force generator--Microtubule attachment modelling}

To account for the limited number of cortical anchors \citep{grill05,pecreaux06}, we modelled the attachment of a force generator to a microtubule  \citep{nguyen07} as a first order process, using the law of mass action on component quantity \citep{koonce12} and combined it to the equations of quantity conservation for force generators and microtubules. It corresponded to the pseudo-chemical reaction: \\ 
\begin{center}
\ce{ Microtubule + Force-generator -> MT\textendash\textendash Force-generator}
\end{center}
and the equilibrium equation came readily:
\begin{subeqnarray}
K_a & = & \frac{N_\text{microtubule--force-generator}}{N_\text{free-microtubule-at-cortex}\,N_\text{free-force-generator}} 
\label{mass-action-number}\\[12pt]
N & = & N_\text{Microtubule--force-generator}+N_\text{free-force-generator} \\[12pt]
M &= & N_\text{microtubule--force-generator}+N_\text{free-microtubule-at-cortex} \\
    & = & \mathcal{M}\left(\mathcal{S},\alpha\right)  \label{conservation_MT}
\end{subeqnarray}
where $N$ is the total number of force generators present in the active region.

We could relate the association constant $K_a$ to our initial model \citep{pecreaux06} (\S \ref{orig}) by writing
\begin{equation}
 K_a= \begin{cases}
 \widehat{k_{on}} \left/ \overline{k_{off}} \right.  \\[10pt]
 \bar{p} \left/ \left(1-\bar{p} \right) \right.\left/N_\text{free-microtubule-at-cortex}\right. \\[10pt]
 \end{cases}
 \label{eq_Ka}
 \end{equation}
with
 the on-rate $k_{on}=\widehat{k_{on}}N_\text{free-microtubule-at-cortex} \label{eq_kon} $,
and 
the off-rate $\overline{k_{off}}(t)$ thought to depend on mitosis progression. Time dependences were omitted for sake of clarity. It was noteworthy that $k_{on}$, used in the initial model as force generator binding rate (assuming microtubules in excess), became variable throughout mitosis in the expanded model as it depends on the number of free microtubule contacts at the cortex, thus on the centrosome position. In contrast, $\widehat{k_{on}}$ appeared constant in the expanded model representing the  on-rate of the first order reaction above.

\paragraph{Related parameter estimate}

In modelling anaphase oscillation onset, we assumed that the off-rate dependence on mitosis progression was negligible (\S \ref{twoSwitches} and \ref{simulPostDisp} for full model without this assumption). The positional switch modelled here led to a limited number of engaged force generators at oscillation onset. At this time, the force generator quantity just crossed the threshold to build oscillations \citep{pecreaux06} and we estimated that typically 70\% of the force generators were thus engaged, consistent with the quick disappearance of oscillations upon progressively depleting the embryo from GPR-1/2 proteins. 
We observed that the oscillation started when the centrosome reached 71\% of embryo length \citep[Table 1]{bouvrais18}. At that moment, 52 microtubules were contacting the cortex (\S \ref{MT_estimates}). We set the total number of force generators to 50 and got a number of engaged ones consistent with previous reports \citep{grill03}. We thus estimated the association constant $K_a^0\simeq 0.1$ (denoted with 0 superscript to indicate that we assumed negligible its variation throughout mitosis). 
In turn, we estimated $\widehat{k_{on}} \simeq 0.025 \;\text{s}^{-1}$ assuming that the detachment rate at that time was about $4\;\text{s}^{-1}$ \citep{rodriguez17}. If 70\% of the force generators were engaged at oscillation onset, it would correspond to $k_{on}\simeq 0.375\;\text{s}^{-1}$, thus comparable to the estimate of this parameter in the initial model \citep{pecreaux06}.

\paragraph{Modelling the number of engaged force generators in the posterior crescent}

In mitosis early stages, when the spindle lays in the middle of the embryo ({ \it C. elegans}) or slightly anteriorly ({\it C. briggsae}), both centrosomes are far from their respective cortex and thus the imbalance in active force generator quantity due to embryo polarity results in a slight posterior pulling force and causes a slow posterior displacement. The closer the posterior centrosome gets to its cortex, the larger the force imbalance (because more microtubules reach the cortex), and the posterior displacement accelerates to (potentially) reach an equilibrium position during metaphase resulting in a plateau in posterior centrosome displacement located around 70\% of the AP axis. Once anaphase is triggered, the decreased coupling between anterior and posterior centrosomes results into a sudden imbalance in favour of posterior pulling forces so that the posterior displacement speeds up \citep{bouvrais18}. 

We quantitatively modelled this phenomenon by combining the law of mass action above (Eq. \ref{mass-action-number}a) with the number of microtubules reaching the posterior crescent (Eq. \ref{Mcal}) to obtain the number of engaged force generators in the posterior cortex as following:
\begin{eqnarray}
\begin{split}
\label{number_eq} 
\mathcal{N} & \left(\mathcal{M}\left(\mathcal{S},\alpha\right)\right) = N \frac{\phi-1}{\phi+1}
 \\
& \text{with} \quad
\phi= \zeta^-+\sqrt{1+{\zeta^-}^2+2\zeta^+}
 \\
 & \zeta^\pm=K_a\left(  \mathcal{M}\left(\mathcal{S},\alpha\right) \pm N \right)
\end{split}
\end{eqnarray}

\manuallabel{ST2}{2}

\begin{figure}
   \begin{center}
      \includegraphics*[width=160mm]{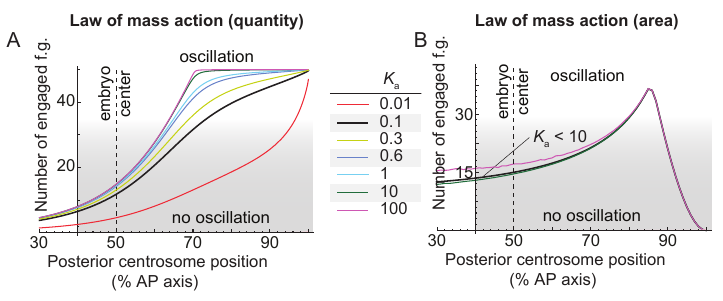}
   \end{center}
      \caption{\textbf{Comparison of mass action law models using quantity and areal concentrations.} When varying the force generator--microtubule association constant $K_a$, graph of the number of engaged force generators (f.g.) versus the posterior displacement of the centrosome along the anteroposterior (AP) axis. For the centrosome positions above 60\% of the AP axis, the number of engaged f.g. steeply increases, saturating above 70\% and creating a switch-like behaviour. Force-generator--microtubule binding was modelled by the law of mass action in: (A) quantity, with total number of force generators $N$ = 50; and (B) areal concentration (\S \ref{sec_Ka}), with $N$ = 500. In both cases, we got similar numbers of engaged force generators between 10 to 100 that are consistent with experimental estimates \citep{grill03} (\S \ref{surcConc}). The parameters used are listed in \citep[Table S5]{bouvrais18}. Grey shading indicates when the engaged force generator count was too low to permit oscillation (below threshold).}
      \label{fig3}
\end{figure}

To challenge our expanded model, we tested the switch-like behaviour in a broad range of association constants $K_a$ (Figure \ref{fig3}A). When the posterior centrosome was between 50\% and 70\% of embryo length, we observed that the number of engaged force generators was increased up to a threshold that enabled oscillations, consistently with \citep{pecreaux06}. When the centrosome was posterior enough, practically above 70\% of AP axis, the number of engaged force generators saturated, suggesting that their dynamics were now the control parameters, as proposed in the initial model during anaphase.  We also observed that a minimal binding constant was needed to reach the threshold number of engaged force generators required for oscillations. Interestingly, above this minimal $K_a$, further increase of the binding constant did not alter significantly the positional switch (Figure \ref{fig3}A). This suggested that this positional switch operates rather independently of the force generator processivity. This will be further discussed below (\S \ref{twoSwitches}). 

\paragraph{The positional switch is independent of the total number of force generators, as soon as this quantity is above a threshold}
As we previously suggested that the total number of force generators should not impact the positional switch \citep{riche13}, we calculated the corresponding prediction in our expanded model \citep[Figure S2B]{bouvrais18} and compared it with experimental prediction \citep[Figure S2C]{bouvrais18}. The good match supports our expanded model. In modelling \textit{gpr-1/2} mutant through the total number of force generators $N$, we followed the common thought that asymmetry of active force generator was due to an increased total number of force generators on the posterior side.

We recently proposed that the asymmetry in active force generators could be an asymmetry of force generator association rate to form the trimeric complex that pulls on microtubules \citep{rodriguez17}. GPR-1/2 presence would increase this on-rate. In our expanded model, a decreased on-rate (through {\it gpr-2} mutant) would result in a decrease association constant $K_a$. Like is the previous case, above a certain threshold of $K_a$, the position at which oscillations were set on was not significantly modified (Figure \ref{fig3}A). In conclusion, independently of the details used to model the force imbalance consequence of the polarity (i.e. the total number or the on-rate), the mild depletion of GPR-1/2 experiment, causing a reduced number of active force generators, supported our expanded model.

\subsubsection{The change of regime in the number of microtubules reaching the cortex versus the centrosome position is independent of detailed embryo shape}
The above results were obtained by assuming an ellipsoidal shape for the embryo (an ellipsoid of revolution around the AP axis, prolate or oblate). We wondered whether a slightly different shape could alter the result. We thus repeated the computation, modelling the embryo shape by a super-ellipsoid of revolution, based on super-ellipses (Lam\' e curves) \citep{edwards1892} defined as:
 \begin{equation}
 \left|\frac{X}{a}\right|^n+\left|\frac{Y}{b}\right|^n=1
 \label{def_super_ellipse}
 \end{equation}
with $a$ and $b$ the half length and width, $n$ the exponents, and $(X,Y,Z)$ the cartesian axes with $X$ along the AP axis (long axis), and positive values towards the posterior side. We obtained a similar switch-like behavior (Figure \ref{fig2}). We concluded that the switch-like behaviour was resistant to changes of the detailed embryo shape and thus we performed the remaining investigations with an ellipsoid shape, for sake of simplicity.

\subsubsection{Sensitivity analysis of the oscillation onset position to embryo geometry and microtubule dynamics}
The expanded model offers a regulation of cortical pulling forces, as revealed by oscillation onset, by the position of the centrosome. We therefore investigated how the shape of the embryo could impact the switch. Indeed, various species of nematode display different long and short axes, resulting in variation of scale and eccentricity \citep{farhadifar15}. In \citep[Figure 4]{bouvrais18}, we reported that embryo length has a reduced impact on the switch. In contrast, the embryo width is more influential over the switch (Figure \ref{fig4}A). It is noteworthy that embryo length undergoes a stronger selection in genetic studies in comparison with embryo width \citep{farhadifar16}.

Then, we investigated the sensitivity of the oscillation onset position to parameters describing embryo shape in a different representation. We found a robustness of the position of oscillation onset versus the eccentricity, i.e. variations in embryo length keeping area constant (Figure \ref{fig4}CD), while embryo scale was more influential (Figure \ref{fig4}B). This is perfectly consistent with the positional control, which measures the distances in units of microtubule dynamics (\S \ref{sensitivity}). Consequently, the position at which oscillation starts is highly dependent on microtubule dynamics (Figure \ref{fig4}E).

\begin{figure}
   \begin{center}
      \includegraphics*[width=160mm]{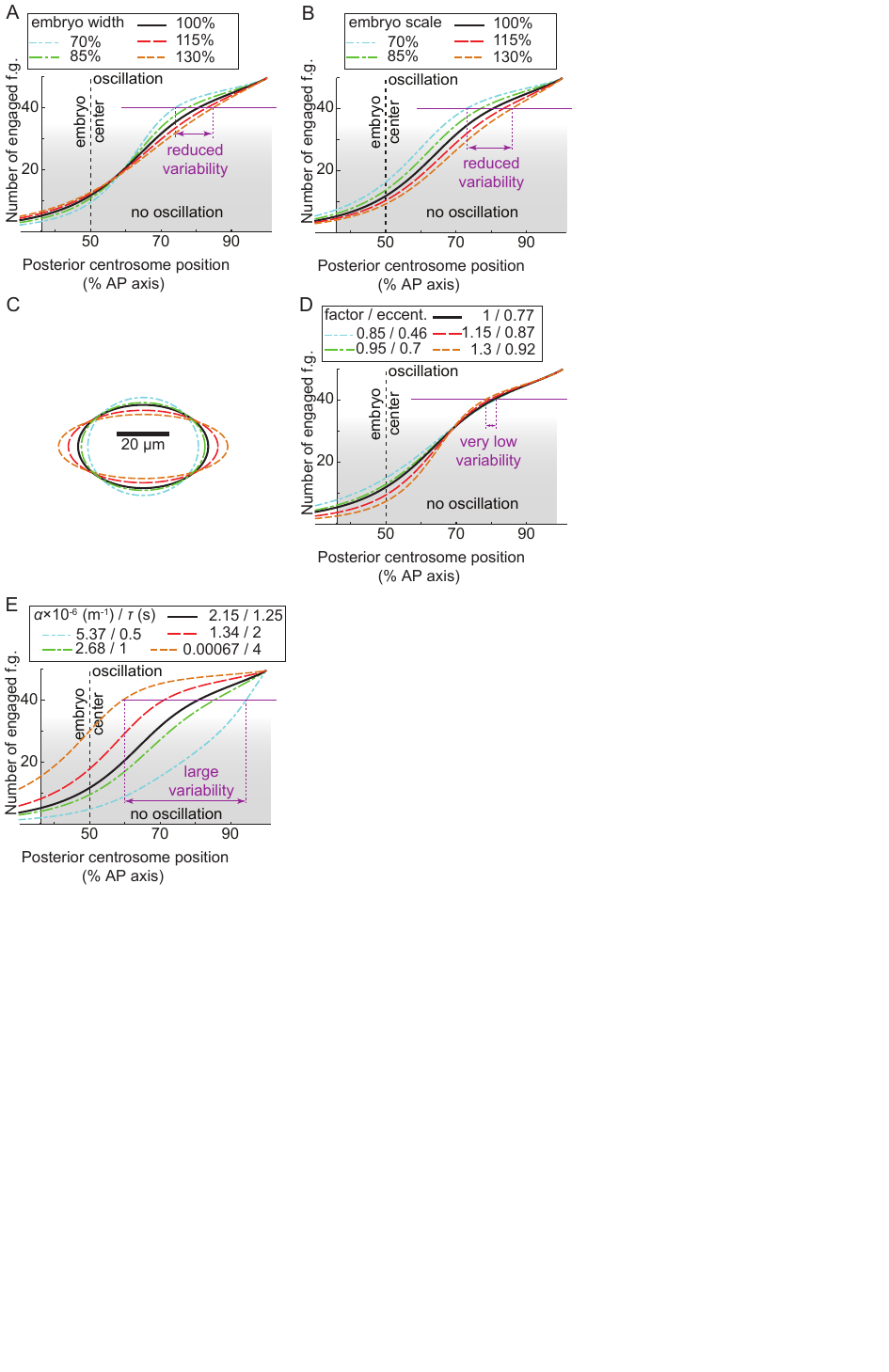}
   \end{center}
   \caption{}
\end{figure}
\begin{figure}
\ContinuedFloat   
      \caption{\textbf{Oscillation onset sensitivity analysis of the expanded model.} Number of engaged force generators (f.g.), versus the posterior displacement of the centrosome along the anteroposterior (AP) axis. (A) Embryo width variations expressed as a percentage of the control. (B) Variations in embryo scale factor on length and width (keeping proportions). (C-D) Variations in the embryo shape after scaling its length by a multiplicative factor and the width by the square root of inverse of this same factor, with the ellipse eccentricity shown in panel C. Doing so, the ellipsoid of revolution modelling the embryo keeps the same volume. (E) Variations in microtubule dynamics summarised by parameter $\alpha$ in $\text{m}^{-1}$ and its equivalent cortical residency time $\tau$ in second, assuming constant growth and shrinkage rates. In all cases, control values are black; green and blue are lower values; and red and orange are higher ones. The parameters used are listed in \citep[Table S5]{bouvrais18}. Grey shading indicates when the number of engaged force generators was too low to permit oscillation (below threshold). Purple thin lines, of equal length in each panel, give a variability scale.}
      \label{fig4}
\end{figure}

\subsubsection{Discussion: number-- or density--limited force generator-microtubule binding \label{surcConc}}
By writing the law of mass action in protein quantity (Eq. \ref{mass-action-number}a), we assumed that the force generator-microtubule binding reaction was rate-limited but not diffusion-limited. We recently investigated the dynamics of cytoplasmic dynein \citep{rodriguez17} and observed that dynein molecules were abundant in cytoplasm, thus 3D diffusion combined to microtubule plus-end accumulation brought enough dynein to the cortex. Therefore, diffusion of dynein to the cortex was not likely to be a limiting factor in binding force generators to the microtubules. However, another member of the force-generating complex, GPR-1/2, essential to generate pulling forces \citep{nguyen07,grill03,pecreaux06}, may be  limiting. GPR-1/2 is likely localised at the cell cortex prior to assembly of the trimeric complex \citep{riche13,park08}, and in low amount, leading to a limited number of cortical anchors \citep{grill03,grill05,pecreaux06}. We thus asked whether a limiting areal concentration of GPR-1/2 at the cortex could alter our model predictions. In the model proposed here, we considered force generator as a reactant of binding reaction. This latter included the molecular motor dynein but also other member of the trimeric complex, as GPR-1/2. Therefore, a limited cortical areal concentration in dynein or GPR-1/2 was modelled identically as a limited areal concentration of force generator. We wrote the corresponding law of mass action in concentration:
 \begin{equation}
\tilde{K}_a = \frac{\left[\text{Microtubule--force-generator}\right]}{\left[\text{Microtubule-at-cortex}\right]\,\left[\text{force-generator}\right]},
\label{mass-action-surfacic}
\end{equation}

with $\left[\text{force-generator}\right]=\frac{N_\text{force-generator}}{S_{active Region}}$, $\tilde{K}_a=K_a  S_{active Region}$ and $S_{active Region}$ the posterior crescent surface (active region), whose boundary is considered at $70\%$ of embryo length. Modelling the embryo by a prolate ellipsoid of radii $24.6\;\mu\text{m}$ and twice $15.75\;\mu\text{m}$, we obtained $S_{active Region}\simeq 0.147\, S_{embryo} = 610\;\mu\text{m}^{-1}$, while the whole embryo surface was $S_{embryo}\simeq4100\;\mu\text{m}^2$.

The probability of a microtubule to hit the cortex (Eq. \ref{eq_p} and \ref{Mcal} ) was modified as follow:
\begin{equation}
\tilde{\mathcal{M}}\left(\mathcal{S},\alpha,\theta,\phi\right) = \frac{M}{4\pi}\frac{1}{1+\alpha\, r_\mathcal{S}\left(\theta,\phi\right)} {\color{blue} \frac{1}{r_\mathcal{S}\left(\theta,\phi\right)^2}}
\label{MT_proba_new}
\end{equation}
We then calculated the number of engaged force generators as above (Eq. \ref{number_eq} ) and found also a positional switch (Figure \ref{fig3}B compared to \ref{fig3}A). We concluded that this alternative modelling of force generator--microtubule attachment was compatible with the positional switch that we observed experimentally. 

In contrast with the law of mass action in quantity, when the centrosome was further displaced towards the posterior after the positional switch, we did not observe any saturation in engaged force generators but a decrease (Figure \ref{fig3}B). This may suggest that the centrosome position could control the oscillation die-down, if diffusion of member(s) of the trimeric complex in the cortex was the limiting factor. In such a case, one would expect that die-down did not intervene after a fixed delay from anaphase onset, but at a given position. This contrasted with experimental observations upon delaying anaphase onset \citep[Table 1]{bouvrais18}. Therefore, the law of mass action in quantity appeared to better model our data.  

On top of this experimental argument, we estimated the lateral diffusion of the limited cortical anchors, likely GPR-1/2, and calculated a corresponding diffusion limited reaction rate equal to $k_{on}^D\simeq 1.2\times 10^5\;\text{s}^{-1}$ after \citep{freeman83a,freeman83b}. We considered the parameters detailed previously, a diffusion coefficient for GPR-1/2 similar to the one of PAR proteins $D=0.2\;\mu\text{m}^2/\text{s}$ \citep{goehring11}, and a hydrodynamic radius of $5.2\;\text{nm}$ \citep{erickson09}. Compared to the on-rate value proposed above (\S \ref{sec_Ka}), i.e. $k_{on}\simeq 0.375 \; \text{s}^{-1}$, this suggested that lateral diffusion was not limiting. In contrast, it was proposed that in such a case, lateral diffusion may even enhance rather than limit the reaction \citep{adam68}. We concluded that the process was limited by reaction, not diffusion, and we considered action mass in quantity (Eq.  \ref{mass-action-number}a) in the remaining of this work. 

\subsubsection{The processivity and microtubule dynamics set two independent switches on force generators: the expanded tug-of-war model \label{twoSwitches}}

We next asked whether a cross-talk exists between the control of the oscillation onset by the processivity, as previously reported \citep{pecreaux06}, and the positional switch explained above. To do so, we let $K_a$ varying with both the processivity $1\left/\overline{k_{off}}\right.$ and 
the centrosome position. In the notations of the initial model, since we kept $\widehat{k_{on}}$ constant, it meant that $k_{on}$ varied because of changes in the number of microtubule contacts in the posterior crescent, in turn depending on the centrosome position. 
We then computed the pairs $(\overline{k_{off}^c},x^c)$ so that Eq. \ref{eqToW} was critical, i.e. $ \Xi^c =\Gamma^t$ (Eq. \ref{eqXi}), with $x^c$ the critical position of the centrosome along the AP axis and $\overline{k_{off}^c}$ the critical off-rate. Because we considered the transverse axis and a single centrosome, we used 
$\Gamma^t=140\;\mu\text{N.s/m}$ after \citep{garzon16} and obtained the diagram reproduced in \citep[Figure 5A]{bouvrais18} that could be seen as a stability diagram. When the embryo trajectory (the orange arrow) crosses the first critical line (collection of $(\overline{k_{off}^c},x^c)$, depicted in blue) to go into the unstable region (blue area), the oscillations start and develop. Since this line is diagonal, it suggests that such an event depends upon the position of the posterior centrosome (ordinate axis) and the detachment rate (abscissa), suggesting that two control parameters contribute to making the system unstable and oscillating. Interestingly, when the embryo continues its trajectory in the phase diagram, it crosses the second critical line (depicted in green), which corresponds to the moment the system becomes stable again, and oscillations are damped out. This critical line is almost vertical indicating that this event depends mostly on the detachment rate, i.e. the inverse of processivity, consistent with the experimental observations \citep[Table 1]{bouvrais18}. Interestingly, this behaviour is maintained despite modest variations in the range of processivity and centrosome position explored during the division (i.e. the precise trajectory of the embryo in this stability diagram). Note that large values of detachment rate are irrelevant as they do not allow posterior displacement of the spindle \citep[Figure 7C, orange curve]{bouvrais18}.  We concluded that two independent switches control the onset of anaphase oscillations and broadly the burst of pulling forces contributing to spindle elongation and posterior displacement.


\section{Simulating posterior displacement and final position \label{simulPostDisp}}
Because the cortical pulling forces involved in the anaphase spindle oscillations are also causing the posterior displacement, and because they depend on the position of the posterior centrosome, it creates a feedback loop on the posterior centrosome position. Resistance to changes of some parameters revealed by the sensitivity analysis of the oscillation onset suggests that these same parameters may have a reduced impact on the final position of the centrosome. In turn, this final position is essential as it contributes to determine the position of the cytokinesis cleavage furrow, a key aspect in an asymmetric division to correctly distribute cell fate determinants \citep{white12,rappaport71,knoblich10}. 

To simulate the kinematics of posterior displacement, we considered the expanded model (\S \ref{EToW}) and a slowly-varying binding constant $K_a$ due to the processivity increasing throughout mitosis (\S \ref{sec_Ka}). We calculated the 
posterior pulling forces, assuming an axisymmetric distribution of force generators. The projection of the force exerted by the cortical pulling force generators implied a weakening factor because only the component parallel to the AP axis contributes to displace posteriorly the spindle. To calculate it, we made the assumption that any microtubule contacting the cortex in the active region has an equal probability to attach a force generator. Therefore, we obtained the force weakening due to AP axis projection by writing the ratio of the forces exerted by each microtubule contacting the cortex weighted by the probability of a contact and integrated over the active region, over the number of microtubule contacts calculated using Eq. \ref{MT_proba_new}. This weakening ratio was then multiplied by the number of bound force generators previously obtained (Eq. \ref{number_eq}). The weakened of the pulling force along AP axis $\mathcal{F}$ then reads:
\begin{equation}
		\mathcal{F}^{ante | post} \left(x,\overline{k_{off}}\right) = \, \frac{ 2\pi \int_{\theta=0}^{\theta_0^{ante | post}}   p\left(\mathcal{S},\alpha, x,\theta \right) \cos\theta\operatorname{d}\theta}{P\left(\mathcal{S},\alpha, x \right)} \times 
		 \mathcal{N} \left(\mathcal{S},\alpha,x, K_a^{ante | post} \right) \bar{f}
   \label{simul_eq_part1} 
\end{equation} 
with $\theta_0$ the polar angle of the active region boundary positioned at $x_{ante}^0$ and $x_{post}^0$, obtained assuming an ellipsoidal shape for the embryo. $p\left(\mathcal{S},\alpha, x_{ante | post},\theta \right)$ was defined at Eq. \ref{eq_p} and $P\left(\mathcal{S},\alpha, x_{ante | post} \right)$ at Eq. \ref{eq_P}. The Eq. \ref{simul_eq_part1} was used to calculate both anterior and posterior forces, with their respective parameters. After \cite{rodriguez17}, the force asymmetry was due to an asymmetry of f.g.-MT affinity, under the control of GPR-1/2. We accounted for this asymmetric on-rate through an asymmetric attachment constant writing $K_a^{ante | post}=\beta^{ante | post} \widehat{kon} \left/ \overline{k_{off}} \right.$.

We put the above quantities into Eq \ref{eqToW} to finally get:
   \begin{equation} \label{simul_eq_part2}
I^{post}\,\ddot{x}_{post}+\left(\Gamma - \Xi^{post}\, \right) \dot{x}_{post} + K\,x_{post}-K_{ante}\,x_{ante} = \eta + \mathcal{F}^{post}(x_{post}) - \mathcal{F}_{ante} 
\end{equation}
with $\eta$ a white noise modelling the force generator stochastic attachment and detachment \citep{pecreaux06,nadrowski04}. In particular, we used \[ k_{on}=\widehat{k_{on}}\left( \mathcal{M}\left(\mathcal{S},\alpha,x_{post}\right) - \mathcal{N} \left(\mathcal{S},\alpha,x_{post},K_a^{post}\right) \right) \] and also applied a weakening of anterior force to account for the uncoupling of spindle poles at anaphase onset \citep{mercat17,maton15}. With $\lambda$ the weakening factor and $\overline{k_{off}^0}$ the force generator off-rate at anaphase onset, we wrote:
     \begin{subeqnarray} \label{simul_eq_part3}
            {F}_{ante}  =
            \begin{cases}
            	\mathcal{F}_{ante} & \text{if} \quad \overline{k_{off}} \geq \overline{k_{off}^0} \\
            	\lambda \mathcal{F}_{ante} & \text{if} \quad \overline{k_{off}}<\overline{k_{off}^0} 
            \end{cases} \\
    \end{subeqnarray} 
    Similarly, the centering force \citep{pecreaux16,garzon16} was also weakened:
           \begin{subeqnarray} \label{simul_eq_part4}
	 K_{ante}  =
            \begin{cases}
            	K & \text{if} \quad \overline{k_{off}} \geq \overline{k_{off}^0} \\
            	\lambda K & \text{if} \quad \overline{k_{off}}<\overline{k_{off}^0} 
            \end{cases}  \\
    \end{subeqnarray} 

\begin{figure}
   \begin{center}
      \includegraphics*[width=160mm]{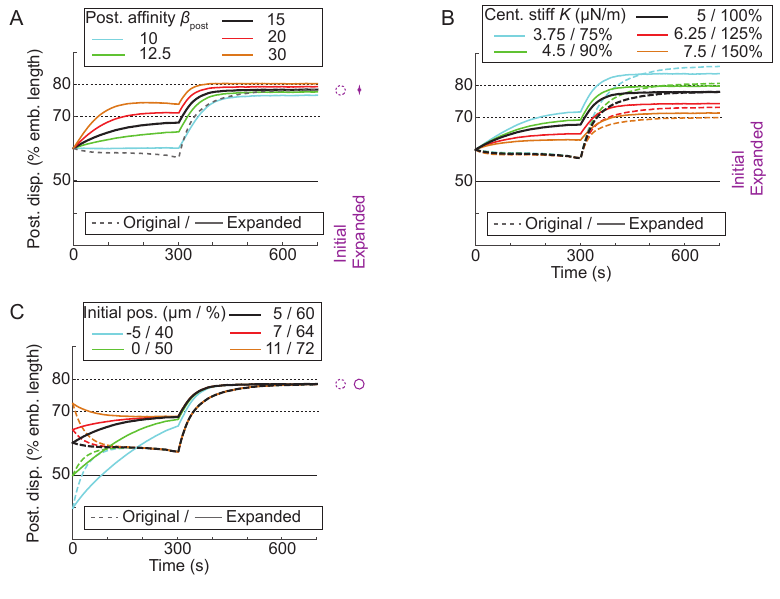}
   \end{center}
      \caption{\textbf{Final position sensitivity analysis of the expanded model.} Stochastic simulations of the posterior centrosome displacement. Dashed lines represent the results of the initial model \citep{pecreaux06}, while solid lines correspond to the expanded one. Posterior displacement of the posterior centrosome averaged over 25 simulation runs with respectively varied: (A) the posterior affinity of the f.g. for microtubule (on-rate) varying $\beta$, whose asymmetry may encode the polarity \citep{rodriguez17}; (B) the centring spring used to model centring forces \citep{pecreaux16} and (C) the initial position of the posterior centrosome. When it does not depend on the parameter considered, the original model is shown by a grey dashed line. In all cases, the control values are black; lower values are blue and green; and the higher values are red and orange. The dispersions of the final values for each case are represented by purple arrows, and a larger span in the plot reveals a lack of robustness to parameter variations. A circle is used when the parameter has no effect on the final value. The parameters used are listed in \citep[Table S5]{bouvrais18}.}
      \label{fig5}
\end{figure}

We solved this system numerically using trapezoidal rule and backward differentiation formula of order 2 (TR-BDF2 algorithm) \citep{hosea96}. Since we linearised the equations and kept the anterior centrosome at a fixed position, we could explore only reasonable parameter variations when performing the final position parameter sensitivity analysis \citep[Figures 6A, 7A-C, 8 ]{bouvrais18}  (Figure \ref{fig5}). As a sanity check, we observed that modest variations in the force generator on-rate, thought to translate polarity cues \citep{rodriguez17}, modulated the final position \citep[Figure 7A]{bouvrais18} as expected from experiments \citep{grill01,colombo03}. To ensure that our simulation correctly converged to the final position, we varied the spindle's initial position and observed no significant change in its final position (Figure \ref{fig5}C).

\section{Conclusion}

We previously proposed that the final centrosome position was dictated both by the centering force stiffness and by the imbalance in pulling force generation, i.e. mainly the active force generator number in active region and their processivity \citep{pecreaux06}. In contrast, in the expanded model, when the posterior centrosome enters into the active region, more microtubules are oriented along the transverse axis than parallel to the AP axis \citep[Figure 9, middle and right panels]{bouvrais18} because of the isotropic distribution of the microtubules around the centrosome. Then, it limits the pulling forces on the posterior centrosome \citep[Figure S3D]{bouvrais18}. As a consequence, the boundary of the active region sets the final position as seen experimentally \citep{krueger10,bouvrais18}. In contrast, the force generator quantity and dynamics become less important and the final position even shows some resistance to changes in these two parameters \citep[Figure 7A-C]{bouvrais18}.

We noticed that when the active region boundary was located at 80\% of embryo length or more posteriorly, and the spindle was close to the cell centre, the number of microtubules reaching this region was so reduced that it prevented a normal posterior displacement. Together with the observation that when the region extended more anteriorly the final position was anteriorly shifted, it appeared that a boundary at 70\%  was a value quite optimal to maximise the posterior displacement. Because this posterior displacement is a key to asymmetric division, it would be interesting (but out of the scope of this work) to see whether a maximal posterior displacement is an evolutive advantage, which would then cause a pressure on the active region boundary. \\

\bibliography{Master}

\begin{thebibliography}{}

\bibitem[Adam and Delbruck, 1968]{adam68}
Adam, G. and Delbruck, M. (1968{\rm{}}).
\newblock Reduction of dimensionality in biological diffusion processes.
\newblock In {\rm Structural Chemistry and Molecular Biology}, (Rich, A.,
  Davidson, N.  and Pauling, L., eds), pp. 198--215. W. H. Freeman San
  Francisco.

\bibitem[Ahringer, 2003]{ahringer03}
Ahringer, J. (2003{\rm{}}).
\newblock Control of cell polarity and mitotic spindle positioning in animal
  cells.
\newblock {\rm Current Opinion in Cell Biology } \emph{15}, 73--81.

\bibitem[Bouvrais {\rm et~al.}, 2018]{bouvrais18}
Bouvrais, H., Chesneau, L., Pastezeur, S., Delattre, M.  and Pecreaux, J.
  (2018{\rm{}}).
\newblock LET-99-dependent spatial restriction of active force generators makes
  spindle's position robust.
\newblock {\rm Biophys J } \emph{submitted}.

\bibitem[Colombo {\rm et~al.}, 2003]{colombo03}
Colombo, K., Grill, S.~W., Kimple, R.~J., Willard, F.~S., Siderovski, D.~P.
  and Gonczy, P. (2003{\rm{}}).
\newblock Translation of polarity cues into asymmetric spindle positioning in
  Caenorhabditis elegans embryos.
\newblock {\rm Science } \emph{300}, 1957--61.

\bibitem[Edwards, 1892]{edwards1892}
Edwards, J. (1892{\rm{}}).
\newblock {\rm An elementary treatise on the differential calculus, with
  applications and numerous examples}.
\newblock Macmillan and Co. The Macmillan Company, London, New York,.

\bibitem[Erickson, 2009]{erickson09}
Erickson, H.~P. (2009{\rm{}}).
\newblock Size and shape of protein molecules at the nanometer level determined
  by sedimentation, gel filtration, and electron microscopy.
\newblock {\rm Biol Proced Online } \emph{11}, 32--51.

\bibitem[Farhadifar {\rm et~al.}, 2015]{farhadifar15}
Farhadifar, R., Baer, C.~F., Valfort, A.~C., Andersen, E.~C., Muller-Reichert,
  T., Delattre, M.  and Needleman, D.~J. (2015{\rm{}}).
\newblock Scaling, selection, and evolutionary dynamics of the mitotic spindle.
\newblock {\rm Curr Biol } \emph{25}, 732--40.

\bibitem[Farhadifar {\rm et~al.}, 2016]{farhadifar16}
Farhadifar, R., Ponciano, J.~M., Andersen, E.~C., Needleman, D.~J.  and Baer,
  C.~F. (2016{\rm{}}).
\newblock Mutation Is a Sufficient and Robust Predictor of Genetic Variation
  for Mitotic Spindle Traits in Caenorhabditis elegans.
\newblock {\rm Genetics } \emph{203}, 1859--70.

\bibitem[Freeman and D., 1983]{freeman83a}
Freeman, D.~L. and D., D.~J. (1983{\rm{}}).
\newblock The influence of diffusion on surface reaction kinetics.
\newblock {\rm The Journal of Chemical Physics } \emph{78}, 6002.

\bibitem[Freeman and Doll, 1983]{freeman83b}
Freeman, D.~L. and Doll, J.~D. (1983{\rm{}}).
\newblock Langevin analysis of the diffusion model for surface chemical
  reactions.
\newblock {\rm The Journal of Chemical Physics } \emph{79}, 2343.

\bibitem[Garzon-Coral {\rm et~al.}, 2016]{garzon16}
Garzon-Coral, C., Fantana, H.~A.  and Howard, J. (2016{\rm{}}).
\newblock A force-generating machinery maintains the spindle at the cell center
  during mitosis.
\newblock {\rm Science } \emph{352}, 1124--7.

\bibitem[Goehring {\rm et~al.}, 2011]{goehring11}
Goehring, N.~W., Hoege, C., Grill, S.~W.  and Hyman, A.~A. (2011{\rm{}}).
\newblock PAR proteins diffuse freely across the anterior-posterior boundary in
  polarized C. elegans embryos.
\newblock {\rm J Cell Biol } \emph{193}, 583--94.

\bibitem[Grill {\rm et~al.}, 2001]{grill01}
Grill, S.~W., Gonczy, P., Stelzer, E.~H.  and Hyman, A.~A. (2001{\rm{}}).
\newblock Polarity controls forces governing asymmetric spindle positioning in
  the Caenorhabditis elegans embryo.
\newblock {\rm Nature } \emph{409}, 630--3.

\bibitem[Grill {\rm et~al.}, 2003]{grill03}
Grill, S.~W., Howard, J., Schaffer, E., Stelzer, E.~H.  and Hyman, A.~A.
  (2003{\rm{}}).
\newblock The distribution of active force generators controls mitotic spindle
  position.
\newblock {\rm Science } \emph{301}, 518--21.

\bibitem[Grill {\rm et~al.}, 2005]{grill05}
Grill, S.~W., Kruse, K.  and Julicher, F. (2005{\rm{}}).
\newblock Theory of mitotic spindle oscillations.
\newblock {\rm Physical Review Letters } \emph{94}, 108104.

\bibitem[Hosea and Shampine, 1996]{hosea96}
Hosea, M.~E. and Shampine, L.~F. (1996{\rm{}}).
\newblock Analysis and implementation of TR-BDF2.
\newblock {\rm Applied Numerical Mathematics } \emph{20}, 21--37.

\bibitem[Howard, 2006]{howard06}
Howard, J. (2006{\rm{}}).
\newblock Elastic and damping forces generated by confined arrays of dynamic
  microtubules.
\newblock {\rm Physical biology } \emph{3}, 54--66.

\bibitem[Kimura and Onami, 2007]{kimura07}
Kimura, A. and Onami, S. (2007{\rm{}}).
\newblock Local cortical pulling-force repression switches centrosomal
  centration and posterior displacement in C. elegans.
\newblock {\rm J Cell Biol } \emph{179}, 1347--54.

\bibitem[Knoblich, 2010]{knoblich10}
Knoblich, J.~A. (2010{\rm{}}).
\newblock Asymmetric cell division: recent developments and their implications
  for tumour biology.
\newblock {\rm Nat Rev Mol Cell Biol } \emph{11}, 849--60.

\bibitem[Koonce and Tikhonenko, 2012]{koonce12}
Koonce, M.~P. and Tikhonenko, I. (2012{\rm{}}).
\newblock Dynein Motor Mechanisms.
\newblock In {\rm Dyneins : structure, biology and disease}, (King, S.~M.,
  ed.), pp. xv, 639 p. Academic Press Amsterdam ; Boston 1st edition.

\bibitem[Kozlowski {\rm et~al.}, 2007]{kozlowski07}
Kozlowski, C., Srayko, M.  and Nedelec, F. (2007{\rm{}}).
\newblock Cortical microtubule contacts position the spindle in C. elegans
  embryos.
\newblock {\rm Cell } \emph{129}, 499--510.

\bibitem[Krueger {\rm et~al.}, 2010]{krueger10}
Krueger, L.~E., Wu, J.~C., Tsou, M.~F.  and Rose, L.~S. (2010{\rm{}}).
\newblock LET-99 inhibits lateral posterior pulling forces during asymmetric
  spindle elongation in C. elegans embryos.
\newblock {\rm J Cell Biol } \emph{189}, 481--95.

\bibitem[Labbe {\rm et~al.}, 2004]{labbe04}
Labbe, J.~C., McCarthy, E.~K.  and Goldstein, B. (2004{\rm{}}).
\newblock The forces that position a mitotic spindle asymmetrically are
  tethered until after the time of spindle assembly.
\newblock {\rm J Cell Biol } \emph{167}, 245--56.

\bibitem[Maton {\rm et~al.}, 2015]{maton15}
Maton, G., Edwards, F., Lacroix, B., Stefanutti, M., Laband, K., Lieury, T.,
  Kim, T., Espeut, J., Canman, J.~C.  and Dumont, J. (2015{\rm{}}).
\newblock Kinetochore components are required for central spindle assembly.
\newblock {\rm Nat Cell Biol } \emph{17}, 697--705.

\bibitem[Mercat {\rm et~al.}, 2017]{mercat17}
Mercat, B., Pinson, X., Le~Cunff, Y., Fouchard, J., Mary, H., Pastezeur, S.,
  Gachet, Y., Tournier, S., Bouvrais, H.  and Pecreaux, J. (2017{\rm{}}).
\newblock Micro-fluctuations of spindle length reveal its dynamics over cell
  division.
\newblock (in preparation).

\bibitem[Nadrowski {\rm et~al.}, 2004]{nadrowski04}
Nadrowski, B., Martin, P.  and Julicher, F. (2004{\rm{}}).
\newblock Active hair-bundle motility harnesses noise to operate near an
  optimum of mechanosensitivity.
\newblock {\rm Proc Natl Acad Sci U S A } \emph{101}, 12195--200.

\bibitem[Nguyen-Ngoc {\rm et~al.}, 2007]{nguyen07}
Nguyen-Ngoc, T., Afshar, K.  and Gonczy, P. (2007{\rm{}}).
\newblock Coupling of cortical dynein and G alpha proteins mediates spindle
  positioning in Caenorhabditis elegans.
\newblock {\rm Nature Cell Biology } \emph{9}, 1294--1302.

\bibitem[O'Toole {\rm et~al.}, 2003]{otoole03}
O'Toole, E.~T., McDonald, K.~L., Mantler, J., McIntosh, J.~R., Hyman, A.~A.
  and Muller-Reichert, T. (2003{\rm{}}).
\newblock Morphologically distinct microtubule ends in the mitotic centrosome
  of Caenorhabditis elegans.
\newblock {\rm J Cell Biol } \emph{163}, 451--6.

\bibitem[Park and Rose, 2008]{park08}
Park, D.~H. and Rose, L.~S. (2008{\rm{}}).
\newblock Dynamic localization of LIN-5 and GPR-1/2 to cortical force
  generation domains during spindle positioning.
\newblock {\rm Developmental Biology } \emph{315}, 42--54.

\bibitem[Pecreaux {\rm et~al.}, 2016]{pecreaux16}
Pecreaux, J., Redemann, S., Alayan, Z., Mercat, B., Pastezeur, S.,
  Garzon-Coral, C., Hyman, A.~A.  and Howard, J. (2016{\rm{}}).
\newblock The Mitotic Spindle in the One-Cell C. elegans Embryo Is Positioned
  with High Precision and Stability.
\newblock {\rm Biophys J } \emph{111}, 1773--1784.

\bibitem[Pecreaux {\rm et~al.}, 2006]{pecreaux06}
Pecreaux, J., Roper, J.~C., Kruse, K., Julicher, F., Hyman, A.~A., Grill, S.~W.
   and Howard, J. (2006{\rm{}}).
\newblock Spindle oscillations during asymmetric cell division require a
  threshold number of active cortical force generators.
\newblock {\rm Current Biology } \emph{16}, 2111--22.

\bibitem[Rappaport, 1971]{rappaport71}
Rappaport, R. (1971{\rm{}}).
\newblock Cytokinesis in animal cells.
\newblock {\rm Int Rev Cytol } \emph{31}, 169--213.

\bibitem[Redemann {\rm et~al.}, 2016]{redemann16}
Redemann, S., Baumgart, J., Lindow, N., Fuerthauer, S., Nazockdast, E., Kratz,
  A., Prohaska, S., Brugues, J., Shelley, M.  and Mueller-Reichert, T.
  (2016{\rm{}}).
\newblock Kinetochore Microtubules indirectly link Chromosomes and Centrosomes
  in C. elegans Mitosis.
\newblock {\rm BioRxiv } \emph{060855}.

\bibitem[Riche {\rm et~al.}, 2013]{riche13}
Riche, S., Zouak, M., Argoul, F., Arneodo, A., Pecreaux, J.  and Delattre, M.
  (2013{\rm{}}).
\newblock Evolutionary comparisons reveal a positional switch for spindle pole
  oscillations in Caenorhabditis embryos.
\newblock {\rm Journal of Cell Biology } \emph{201}, 653--62.

\bibitem[Rodriguez~Garcia {\rm et~al.}, 2017]{rodriguez17}
Rodriguez~Garcia, R., Chesneau, L., Pastezeur, S., Roul, J., Tramier, M.  and
  Pecreaux, J. (2017{\rm{}}).
\newblock Dynein dynamics at the microtubule plus-ends and cortex during
  division in the C. elegans zygote.
\newblock {\rm bioRxiv } \emph{}, 118604.

\bibitem[Srayko {\rm et~al.}, 2005]{srayko05}
Srayko, M., Kaya, A., Stamford, J.  and Hyman, A.~A. (2005{\rm{}}).
\newblock Identification and characterization of factors required for
  microtubule growth and nucleation in the early C. elegans embryo.
\newblock {\rm Dev Cell } \emph{9}, 223--36.

\bibitem[White and Glotzer, 2012]{white12}
White, E.~A. and Glotzer, M. (2012{\rm{}}).
\newblock Centralspindlin: At the heart of cytokinesis.
\newblock {\rm Cytoskeleton (Hoboken) } \emph{69}, 882--92.

\end{thebibliography}

\end{document}